\documentclass[12pt,preprint]{aastex}

\usepackage{natbib}
\newcommand{\mps}{{\rm\,m\,s}^{-1}}

\makeatletter
\newcommand*{\rom}[1]{\expandafter\@slowromancap\romannumeral #1@}
\makeatother
\begin{document}

\title{Characterizing the Orbital and Dynamical State of the HD 82943 Planetary System With Keck Radial Velocity Data\footnote{
       Based on observations obtained at the W. M. Keck Observatory,
       which is operated jointly by the University of California and
       the California Institute of Technology.}
}
\author{Xianyu Tan\altaffilmark{2,3}, Matthew J. Payne\altaffilmark{4,5}, Man Hoi Lee\altaffilmark{2,6}, 
       Eric B. Ford\altaffilmark{4}, Andrew W. Howard\altaffilmark{7}, John. A. Johnson\altaffilmark{8}, Geoff W. Marcy\altaffilmark{9}, Jason T. Wright\altaffilmark{10,11}}

\altaffiltext{2}{Department of Earth Sciences, The University of Hong Kong, Pokfulam Road, Hong Kong}
\altaffiltext{3}{Department of Planetary Sciences and Lunar and Planetary Laboratory, The University of Arizona, 1629 University Blvd., Tucson, AZ 85721, USA; xianyut@lpl.arizona.edu}
\altaffiltext{4}{Department of Astronomy, University of Florida, FL 32611-2055, USA}  
\altaffiltext{5}{Harvard-Smithsonian Center for Astrophysics, 60 Garden Street, Cambridge, MA 02138, USA}  
\altaffiltext{6}{Department of Physics, The University of Hong Kong, Pokfulam Road, Hong Kong}
\altaffiltext{7}{Institute for Astronomy, University of Hawaii at Manoa, 2680 Woodlawn Drive, Honolulu, HI 96822, USA}
\altaffiltext{8}{Department of Planetary Sciences,  California Institute of Technology, 1200 E. California Blvd, Pasadena, CA 91125, USA}      
\altaffiltext{9}{Department of Astronomy, University of California, Berkeley, CA 94720-3411, USA}
\altaffiltext{10}{Department of Astronomy and Astrophysics, Pennsylvania State University, University Park, PA 16802, USA}      
\altaffiltext{11}{Center for Exoplanets and Habitable Worlds, The Pennsylvania State
  University, University Park, PA 16802, USA}   
                 
\begin{abstract}
We present  an updated analysis of radial velocity data of the HD
82943 planetary system based on 10 years of measurements obtained with
the Keck telescope. Previous studies have shown that the HD 82943
system has two planets that are likely in 2:1 mean-motion resonance
(MMR), with the orbital periods about 220 and 440 days
\citep{Lee2006}. However, alternative fits that are qualitatively
different have also been suggested, with two planets in a 1:1
resonance \citep{Gozdziewski2006} or three planets in a Laplace 4:2:1
resonance \citep{Beauge2008}. Here we use  $\chi^2$ minimization
combined with parameter grid search  to investigate the orbital
parameters and dynamical states of the qualitatively different types
of fits, and we compare the results to those obtained with the
differential evolution Markov chain Monte Carlo method. Our results
support the coplanar 2:1 MMR configuration for the HD 82943 system,
and show no evidence for either the 1:1 or 3-planet Laplace resonance
fits. The inclination of the system with respect to the sky plane is
well constrained at $20^{+4.9}_{-5.5}$ degrees, and the system contains two
planets with masses of about 4.78 $M_{\rm{J}}$ and 4.80 $M_{\rm{J}}$
(where $M_{\rm{J}}$ is the mass of Jupiter) and orbital periods of
about 219 and 442 days for the inner and outer planet, respectively.
The best fit is dynamically stable with both eccentricity-type
resonant angles $\theta_1$ and $\theta_2 $ librating around
0$^{\circ}$.
\end{abstract}

\section{INTRODUCTION}

To date, more than 125 extrasolar multiple-planet systems have been confirmed, in which there are nearly 40 systems that are suspected to contain planets in or near mean-motion resonance (MMR) (see, e.g., \citealp{Wright11}, \citealp{exoplanet.org}\footnote{See http://exoplanets.org for an up-to-date website.}), hinting that MMRs play an important role in the orbital configurations of  exoplanetary systems. Candidates from the \textit{Kepler} transit survey show a significant fraction of adjacent planets with their period ratio in or near first order mean-motion resonance commensurabilities \citep{Lissauer2011,Fabrycky2012}. However, \citet{Veras2012} have analytically shown that the vast majority of  \textit{Kepler} systems with two near-resonance transiting candidates cannot be in resonances. As such, the fact that the orbital period ratio is close to a low-order integer ratio does not necessarily indicate that a system is in resonance, and that a more detailed examination of the system dynamics is required to rule in or out this possibility.  

The first pair of exoplanets discovered to be in mean-motion resonance was the GJ 876 planetary system \citep{Marcy2001}. By fitting radial velocity (RV) measurements, it is well established that the pair of planets is in a deep 2:1 MMR with both lowest order, eccentricity-type MMR angles,
\begin{equation}
% \begin{split}
      \theta_1 = \lambda_1 - 2\lambda_2 + \varpi_1  , \\
 \end{equation}
 \begin{equation}
      \theta_2 = \lambda_1 - 2\lambda_2 + \varpi_2  ,\\ 
%      \end{split} \label{rangle}
\end{equation}
and the secular apsidal resonance angle,
\begin{equation}
\theta_{\rm{SAR}} = \varpi_1 - \varpi_2 = \theta_1 - \theta_2,
\end{equation}
librating about $0^{\circ}$ with small libration amplitudes \citep{Laughlin2001,Rivera2001,Laughlin2005}.  Here, $\lambda$ is the mean longitude, $\varpi$ is the longitude of periapse, and subscripts 1 and 2 represent the inner and outer planets, respectively. An updated RV analysis by \citet{Rivera2010} showed that the GJ 876 planetary system contains an additional outer planet, and it is in a Laplace resonance with the previously  known 2:1 resonant pair. The orbital configuration of the resonant pair of planets b and c in GJ 876 system differs from that of the Galilean satellites Io and Europa, which is also in 2:1 MMR but with $\theta_1$ librating around $0^{\circ}$ and both $\theta_2$ and  $\theta_{\rm{SAR}}$ librating around $180^{\circ}$. The anti-aligned corotational configuration ensures that  Io and Europa remain stable in the 2:1 MMR because of their small eccentricities \citep{Lee2002}, whereas moderate and large eccentricities allow a wide variety of other stable 2:1 MMR configurations, including the GJ 876 configuration \citep{Lee2002,Beauge2003,Ferraz-Mello2003,Lee2004,Beauge2006}. 
The 2:1 MMR configuration of the GJ 876 system can be established by convergent orbital migration caused by disk-planet interactions, and thus provided constraints on the migration processes within, and the physical environment constituting, the protoplanetary disk. $N$-body simulations with forced inward orbital migration of the outer planet for the GJ 876 system show that either significant eccentricity damping during migration or quick dispersal of  the disk after resonance capture is necessary  to explain the observed eccentricity values \citep{Lee2002}. Although earlier hydrodynamic simulations did not show significant eccentricity damping from disk-planet interactions \citep{Papaloizou2003,Kley2004,Kley2005,Moorhead2008}, a recent study for the GJ 876 system by \citet{Crida2008} shows that  a stronger eccentricity damping from disk-planet interactions can be produced if the inner disk and its interactions with the inner planet are considered. 

It is important to increase the sample of well-determined resonant systems, as the GJ 876 system has shown interesting properties in terms of orbital dynamics as well as providing constraints on the evolution of planetary system. This paper focuses on characterizing the orbital and dynamical state of the  suspected 2:1 MMR system  HD 82943.
The detection of the HD 82943 planetary system was announced in European Southern Observatory (ESO) press releases\footnote{See http://www.exoplanets.ch/hd82943syst.html.}, with the  first planet discovered in 2000 and the second planet in 2001. Based on the orbital parameters posted in the Geneva group web page, \citet{Ji2003} claimed that the fit tends to be unstable unless the system is locked in 2:1 MMR. \citet{Mayor2004} published a double-Keplerian fit of the HD 82943 system based on 142 CORALIE RV measurements, and claimed  that this system is in 2:1 MMR with orbital periods of about 220 and 440 days, and planetary mass ratio of $\sim 1.0$. However, the actual data set was never published. 

By direct $N$-body integrations, \citet{Ferraz-Mello2005} found that the solution of the HD 82943 system found by \citet{Mayor2004} was unstable on the order of $10^5$ yr. The instability of this solution did not change when the integration was started at different initial times instead of starting only at the first observational time epoch of \citet{Mayor2004}, but with the other Keplerian orbital elements the same.  Using the CORALIE RV data extracted from the graphs in \citet{Mayor2004}, \citet{Ferraz-Mello2005} found a best fit with rms = $6.8 \mps$ based on a double-Keplerian model, and explored the  parameter space around the best fit. The best fit was unstable, but the rms as a function of the primary parameters was shallow around the minimum and there were many stable good fits  giving  slightly higher rms. In a statistical sense the real solution may correspond to one of the many stable good fits.  

With 23 additional new RV data from the Keck Observatory, \citet{Lee2006} analyzed the combination of the CORALIE data set (which was extracted from graphs in \citealt{Mayor2004}) and the Keck data set,  using the double-Keplerian model combined with parameter grid search. The best fit had $\chi^2_{\nu}=1.87$ and rms = $7.88 \mps$, but it was also dynamically unstable. Fits in the parameter grid as a function of the eccentricity and argument of periapse of the outer planet were systematically explored. The $\chi^2_{\nu}$ minimum was shallow around the best fit, with good fits that have just slightly higher $\chi^2_{\nu}$ being stable. Dynamical stability exploration of fits in the parameter grid showed that all fits that are stable are in 2:1 MMR, assuming that the planets  are in coplanar orbits. The results suggested that the HD 82943 system was almost certainly in 2:1 MMR.

Using the same data sets, \citet{Gozdziewski2006} undertook a dynamical fitting analysis  using a hybrid algorithm which introduces an orbital instability penalty factor into the $\chi^2_{\nu}$ minimization. The  edge-on coplanar 2:1 MMR best fit with approximately the same $\chi^2_{\nu}$ as the best fit found by \citet{Lee2006} was unstable, but two islands of stable 2:1 MMR fits near the best fit in parameter space were found. Using a  genetic  search algorithm, qualitatively different fits associated with the 1:1 eccentric resonance configuration were found. The best stable fit with a 1:1 resonance  is highly mutually inclined, and just slightly worse than the best fit of 2:1 MMR in terms of $\chi^2_{\nu}$. This raises the interesting possibility that similar radial velocities could be produced by 2:1 MMR and 1:1 eccentric resonance orbits for the HD 82943 system.

Based on the same data sets, \citet{Beauge2008} analyzed the sensitivity of the fits with respect to data set by the so-called Jackknife method. The parameters of the best fit are sensitive to a few data points in the Keck data set, suggesting a plausible inconsistency between the CORALIE and Keck data set.  A new type of fit corresponding to a coplanar 3-planet orbital configuration which included an additional outer planet was suggested. \citet{Beauge2008} claimed that the quality of the edge-on coplanar 3-planet dynamical best fit was a significant improvement accounting for the increased number of fitting parameters. The best-fit  3-planet configuration was unstable, but interestingly, a fit that corresponds to a stable Laplace resonance configuration with $\chi^2_{\nu} = 1.73$ was suggested.

Two major problems remain in the orbital characterization of the HD 82943 planetary system. First, the minimum of $\chi^2_{\nu}$ and rms of the 2:1 MMR best fit was very shallow in parameter space and  almost all of the best fits found to date have been dynamically unstable. Second, based on the combined RV data of CORALIE and Keck observations, three qualitatively different type of fits, i.e., 2:1 MMR fits, 3-planet fits and 1:1 resonance fits, with almost equally good quality have been suggested, complicating the situation on this system.  The purpose of this analysis of the HD 82943 system is to see whether we can discriminate between these three different type of fits, and whether our $N$-body dynamical fit can constrain the inclinations and masses of the planets with the longer time span of the Keck data. In our analysis, we only use the Keck data instead of combining with the CORALIE data for fitting for the following reasons: the CORALIE data were never published  except in graphical form, and the analysis by \citet{Beauge2008} implied an inconsistency between the CORALIE and Keck RV measurements. Although the Keck data is more accurate and has a longer time span, the large number of CORALIE data points means that they could strongly influence the properties of the fits if they are included.  

This paper is organized as follows: In \S 2, we introduce the stellar properties of HD 82943 and the available RV data from observations at Keck Observatory. In \S 3, we introduce our $N$-body dynamical fitting method based the framework of $\chi^2$ minimization. In \S 4, we present our  fitting results which are organized for discriminating three qualitatively different type of fits described above. We show that the coplanar 2:1 MMR configuration is preferred, and the inclination of the system with respect to the sky plane is well constrained at about $20^{\circ}$. In \S 5, distributions and uncertainties of the best-fit parameters are presented. In \S 6, we compare results from $\chi^2$ minimization combined with grid search to those from the Markov chain Monte Carlo analysis.  Finally, our conclusions are summarized in \S 7.

\section{STELLAR CHARACTERISTICS AND RADIAL VELOCITY MEASUREMENTS}

The stellar properties of the G0 star HD 82943 were summarized by \citet{Mayor2004}, \citet{Fischer2005} and \citet{Takeda2007}.
 A stellar mass of $m_0 = 1.18M_{\odot}$ for HD 82943 was suggested by \citet{Laws2003} and \citet{Fischer2005} independently, and a smaller stellar mass of  $m_0 = 1.08M_{\odot}$ was determined by  \citet{Santos2000}. A median mass of $m_0 = 1.15 M_{\odot}$ was adopted by \citet{Mayor2004}, \citet{Lee2006} and \citet{Gozdziewski2006}.
  Here we adopt the slightly lower stellar mass of $m_0 = 1.13 M_{\odot}$ from the more recent  spectroscopic analysis by \citet{Takeda2007}. The age of the star HD 82943 is about 3 Gyr \citep{Mayor2004,Takeda2007}. Thus any good fit that interprets the configuration of the HD 82943 planetary system should be constrained to be dynamically  stable on timescales of the order of $10^9$ yr. Uncertainties in the stellar radial velocity may arise from the stellar jitter caused by acoustic $p$-modes, turbulent convection, star spots, and flows controlled by surface magnetic fields. We adopt a stellar jitter of $4.2\mps$  estimated by \citet{Lee2006} for HD 82943, and the jitter is quadratically added to the internal uncertainties of the RV data sets for results in \S 4 and \S 5. The jitter is treated as an unknown parameter in the Bayesian analysis of \S 6.

We began RV measurements for HD 82943 in 2001 with the HIRES echelle
spectrograph on the Keck \rom{1} telescope. We have to date obtained
64 RV measurements spanning about 10 yr, and the measurements  are
listed in Table \ref{tab:RV}. This is a significant improvement on the
23 Keck measurements spanning 3.8 yr used by \citet{Lee2006}. For the
23 data points previously published by \cite{Lee2006}, the RV in Table
\ref{tab:RV} are slightly different and the uncertainties are smaller,
because we have analyzed all spectra uniformly, using an improved
Doppler analysis code and a new template taken in 2009 with higher
resolution and higher signal-to-noise ratio. The split in the data in
Table \ref{tab:RV} corresponds to the 2004 upgrade of the Keck-HIRES
CCD System. The median of the internal velocity uncertainties for the
two sets of Keck data are about 1.5 and 1.2 $\mps$,
respectively. Because the Keck velocities were measured with two
different instrumental configurations, we model the two different Keck
data sets as having two unknown velocity  offsets.

\section{METHODOLOGY}

More than 8 orbital periods of the outer planet of the HD 82943 planetary system have been observed, and the mutual gravitational interactions between the planets are expected to significantly influence their orbital configuration. Therefore, it is important to adopt a self-consistent dynamical analysis for the RV data. 

The Levenberg-Marquardt (LM) method \citep{Press1992} can efficiently search for the local $\chi^2$ minimum given an initial guess of a set of fitting parameters. A key ingredient in the LM method is the partial derivative of the radial velocity function with respect to each of the fitting parameters $\partial{u}/\partial{\mathbf{a}}$, where $u$ is the radial velocity function and $\mathbf{a}$ is the set of fitting parameters. 
A method for calculating these derivatives in dynamical fitting has been developed from the variational equations that describe the difference of two adjacent orbital motions and its evolution with time. A detailed description of this method and its application to dynamical fitting for multiple planetary systems is given by \citet{Pal2010}. Compared to the numerical derivatives method which has been widely used in dynamical RV fitting, the advantage of the variational equation method is the better accuracy in calculating the derivatives $\partial{u}/\partial{\mathbf{a}}$ and the lower computational cost when the number of planets is not large. 

For coplanar orbits, the fitting parameters are the zero-point offset velocity $V_l$ for each RV data set, the inclination of the system's orbital plane relative to the sky plane $i$ and the  initial osculating  Keplerian orbital elements for each planet, i.e., ($K$, $P$, $e$, $\omega$, $\mathcal{M}$). For mutually inclined orbits, the fitting parameters are the zero-point offset velocity $V_l$ for each RV data set and  the initial osculating  Keplerian orbital elements for each planet, i.e., ($K$, $P$, $e$, $\omega$, $\mathcal{M}$, $i$, $\Omega$). Here, $K$ is the amplitude of the radial velocity, $P$ is the orbital period, $e$ is the orbital eccentricity, $\omega$ is the argument of periapse, $\mathcal{M}$ is the mean anomaly at the first observational epoch (which is BJD 2452006.913 for the current data set), $i$ is the orbital inclination relative to the sky plane and $\Omega$ is the ascending node. Since RV fitting can only determine the difference in the ascending nodes of orbits, not the absolute ascending node for each orbit measured from arbitrary reference direction in the sky plane, it is convenient to fix the initial ascending node of the first planet $\Omega_1$ to  0, and the ascending nodes of the other planets are simply treated as the mutual ascending nodes relative to the first planet at the first observational epoch. Note that during the $N$-body integration in our dynamical fitting, all the orbital parameters are allowed to evolve, so the values of all $\Omega$ (including $\Omega_1$) evolve with time.  In some situations  variants of the fitting parameters are preferred. For example, when the eccentricities are small, we prefer ($h=e\sin \omega$, $k=e\cos \omega$, $\lambda=\omega+\mathcal{M}$) rather than ($e$, $\omega$, $\mathcal{M}$). 
The observed radial velocity $u$ of the central body  is 
\begin{equation}
u = \sum\limits_{j=1}^N -\frac{m_j}{M_{\rm{total}}} (o_x \dot{x}_j + o_y \dot{y}_j + o_z \dot{z}_j) + \sum\limits_{l=1}^L V_l,  \label{rvdyn2}
\end{equation}
where $M_{\rm{total}}=\sum\nolimits_{j=0}^N m_j$ is the total mass of the system, $m_j$ is planetary mass, $(\dot{x}_j, \dot{y}_j, \dot{z}_j)$ are velocity components for planet $j$ in astrocentric coordinates, $N$ is the number of planets, $L$ is the number of data sets, the unit vector $(o_x, o_y, o_z)$ is directed  towards the observer and here we set it to be $(0, 0,1)$. 

The equations of planetary motion in astrocentric coordinates are
\begin{equation}
\dot{m}_j=0   \label{motion1} .
\end{equation}
\begin{equation}
\dot{\mathbf{r}}_j =\mathbf{v}_j   \label{motion2},
\end{equation}
\begin{equation}
\dot{\mathbf{v}}_j =-\frac{G(m_0+m_j)\mathbf{r}_j}{r_j^3}+  \sum\limits_{k=1,k\neq j}^N Gm_k\left(\frac{\mathbf{r}_k-\mathbf{r}_j}{|\mathbf{r}_k-\mathbf{r}_j|^3}-\frac{\mathbf{r}_k}{r_k^3}\right)   .\label{motion3}
\end{equation}
Here, $m_0$ is the mass of the central body, $G$ is the gravitational constant, for planet $j$,  $\mathbf{r}_j$ is the position vector and $\mathbf{v}_j$ is the velocity vector.
 Let us denote the left hand sides of  Eqs. (\ref{motion1}) -- (\ref{motion3}) as variables $\dot{\mathbf{X}}=(\dot{m}, \dot{\mathbf{r}}, \dot{\mathbf{v}})$, and  the right hand sides as functions $F(\mathbf{X})$. The variations $\delta \mathbf{X}$ denote  arbitrary deviations from the original variables $\mathbf{X}$, and the evolution of these deviations with time can be calculated by the linearized variational equations
\begin{equation}
\delta\dot{X}_l = \sum\limits_{m=1}^M \delta{X}_m \cdot\frac{\partial F_l(\mathbf{X})}{\partial X_m} ,\label{vari1}
\end{equation}
where $M$ is the number of parameters. Because of the linear property, the relation between initial variations and variations at arbitrary time $t$ can be written in the form
\begin{equation}
\delta X_k(t)=Z_{kl}\cdot\delta X_l^0   . \label{vari2}
\end{equation}
The matrix $\mathbf{Z}$ measures the ratio of the deviations at an arbitrary time, $t$, to the initial deviations,
and the initial matrix $\mathbf{Z}$ (at $t=0$) is a unit matrix, i.e., $Z_{kl}(t=0)=\delta_{kl}$.
Combining Eqs. (\ref{vari1}) and (\ref{vari2}), the variational equations can be written as  a set of linear differential equations in matrix form
\begin{equation}
\dot{Z}_{kl}=\sum\limits_{m=1}^M Z_{ml}\cdot\frac{\partial F_k(\mathbf{X})}{\partial X_m} \label{vari3}.
\end{equation}
Explicit expressions of the derivatives $\partial{F_k(\mathbf{X})}/\partial{X_m}$ in Eq. (\ref{vari3}) can be found in \citet{Baluev2011}.  
The radial velocity in a dynamical model is the stellar velocity caused by the planets. The partial derivatives of the radial velocity $u(t)$ at time $t$ with respect to initial orbital parameters $\mathbf{a}$  are in fact related to gradients of orbital motions with respect to initial orbital parameters.  These gradients are calculated from the matrix $\mathbf{Z}$ in the variational equations. Thus, the partial derivatives of the radial velocity can be calculated using the chain rule
\begin{equation}
\frac{\partial u(t)}{\partial \mathbf{a}} = \frac{\partial u(t)}{\partial \mathbf{X}(t)}\cdot \mathbf{Z} \cdot\frac{\partial \mathbf{X}^0}{\partial \mathbf{a}}  \label{dyn}.
\end{equation}
Explicit expression for the gradients of $\partial \mathbf{X}^0/\partial \mathbf{a}$ for coplanar orbits with $(h, k, \lambda)$ variables can be found in \citet{Pal2010}.

The initial orbital parameters in hierarchical multiple-planet systems should be interpreted as Jacobi coordinates. The Jacobi coordinates better emulate the assumptions used to fit the parameters to the data than either barycentric coordinates or astrocentric coordinates \citep{Lissauer2001, Lee2003}. The fitting parameters in Jacobi coordinates are denoted as $\tilde{\mathbf{a}}$, and their definition can be found in \citet{Lee2006}. 
In this case, Eq. (\ref{dyn}) should be replaced by
\begin{equation}
\frac{\partial u(t)}{\partial \tilde{\mathbf{a}}} = \frac{\partial u(t)}{\partial \mathbf{X}(t)}\cdot \mathbf{Z} \cdot\frac{\partial \mathbf{X}^0}{\partial \tilde{\mathbf{X}}^0} \cdot \frac{\partial \tilde{\mathbf{X}}^0}{\partial \tilde{\mathbf{a}}}  \label{dyn2} ,
\end{equation}
where $\tilde{\mathbf{X}}^0$ are the variables $(\dot{\tilde{m}}, \dot{\tilde{\mathbf{r}}}, \dot{\tilde{\mathbf{v}}})$ in Jacobi coordinates at the first observed epoch. The transformation from variables $\mathbf{X}$  to $\tilde{\mathbf{X}}$  can be found in \citet{Saha1994}, and it can be written in matrix form
\begin{equation}
\mathbf{X}=\mathbf{A}\cdot\tilde{\mathbf{X}}.
\end{equation}
Thus, the component $\partial \mathbf{X}^0/\partial \tilde{\mathbf{X}}^0$ in Eq. (\ref{dyn2}) is substituted with matrix $\mathbf{A}$.

The variational equations (Eq. [\ref{vari3}]) are integrated simultaneously with the equations of motions (Eqs. [\ref{motion1}]-[\ref{motion3}]) using the Bulirsch-Stoer integrator, and the RV values in the model and their partial derivatives with respect to the fitting parameters (Eqs. [\ref{dyn}] and [\ref{dyn2}]) are obtained in every observational time. There are several versions of the Bulirsch-Stoer integrator, and we adopt the one implemented in the SWIFT package \citep{Levison1994}. Because the LM method is a local $\chi^2$ minimization method, we use the systematic grid search techniques \citep{Lee2006} to ensure the search for global best fit.

\section{FITTING RESULTS}

The HD 82943 planetary system has been studied for years, and previous RV fitting results provide  good initial guesses for $\chi^2$ minimization in our study. However, we must keep in mind that  because our RV data sets are different from previous ones (see the discussion in section 1), other solutions may be found. We first fit with Keplerian model, and the best-fit Keplerian model is taken as an initial guess for dynamical fitting.

\subsection{2:1 MMR Fits} 
\subsubsection{Coplanar Edge-on Fits}

We first conduct a brief Keplerian analysis with the initial parameters adopted from the global best fit (Fit \rom{1}) found by \citet{Lee2006}. The $\chi^2$ reaches a minimum, similar to that of Fit \rom{1} in the upper left corner of the $e_2-\omega_2$ grid shown in Fig. 3 of \citet{Lee2006}. We then explore fits in the $e_2-\omega_2$ grid starting from this minimum.  There is a second minimum in the lower left corner of the $e_2-\omega_2$ grid lower in $\chi^2_{\nu}$. Based on the lower minimum, we allow all parameters to vary and obtain the best fit with $\chi^2_{\nu} = 1.38$ and $\rm{rms}=4.66\mps$. The parameters of the best fit is listed in the first part of Table \ref{tab:best21}. The $\chi^2_{\nu}$ and rms are significantly improved compared to Fit \rom{1} of \citet{Lee2006} ($\chi^2_{\nu} = 1.87, \rm{rms}=7.88\mps$), and the Keplerian best fit is dynamically stable for at least $10^8$ yr according to a direct $N$-body integration.

\begin{figure}
\centering 
\includegraphics[width=0.95\textwidth]{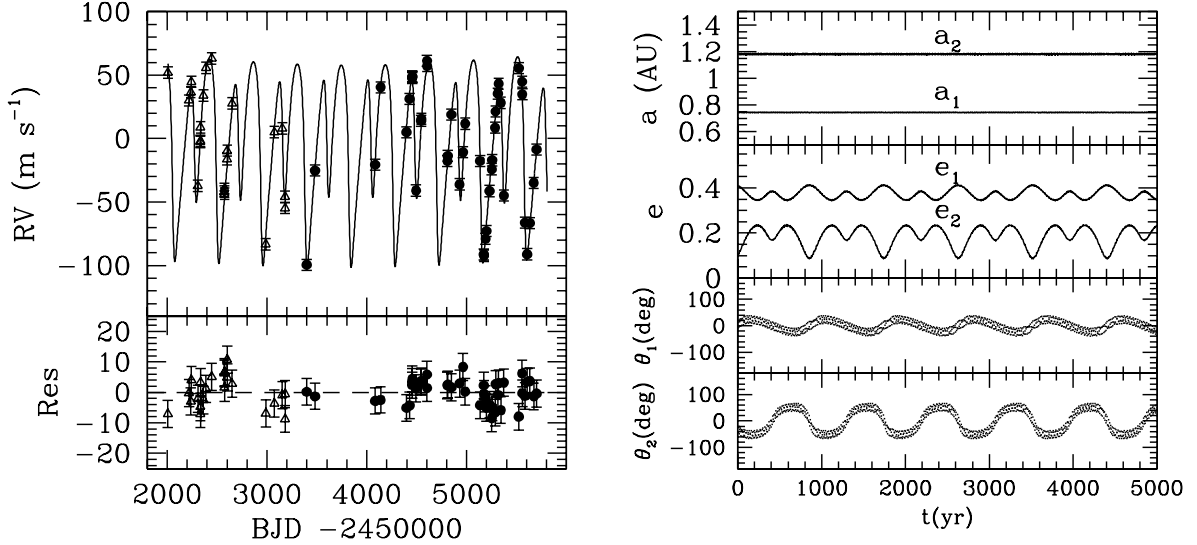}
\caption{  Left panel shows the RV curve and residuals of the 2:1 MMR coplanar edge-on \emph{dynamically interacting} best fit. Open triangles are data points from data set 1, and filled circles are data points from data set 2 (see Table \ref{tab:RV}). Right panel shows semimajor axes $a_1,a_2$, eccentricities $e_1,e_2$ and resonance angles $\theta_1=\lambda_1-2\lambda_2+\varpi_1,~ \theta_2=\lambda_1-2\lambda_2+\varpi_2$ from dynamical evolution of the best fit (they are interpreted in \emph{Jacobi} coordinates). At the epoch of the observation, $e_2$ is small ($\sim 0.09$), but over $\sim$ 1,000 yrs fluctuates significantly due to the large libration amplitudes of $\theta_1, \theta_2$.
\label{fig:RV2190}
}
\end{figure}

The Keplerian best fit is used as an initial guess for coplanar edge-on dynamical fitting. The LM algorithm converges to a minimum with  $\chi^2_{\nu} = 1.21$ and $\rm{rms} = 4.37\mps$.  Starting from different initial guesses, we do not find any other local minimum, so this is likely  the global best fit for the coplanar edge-on dynamical case. The parameters of the best fit are listed in the second part of Table \ref{tab:best21} and the RV curve and residuals are shown in the left panel of Fig. \ref{fig:RV2190}. The dynamical fitting is an improvement over the Keplerian fitting, as the $\chi^2$ of the coplanar edge-on dynamical best fit is less than the $\chi^2$ of the Keplerian best fit by about 8.8.  The eccentricity  of the outer planet ($e_2=0.096$) is smaller than the best fit ($e_2=0.219$) found by \citet{Lee2006}. The edge-on best fit is dynamically stable for at least $\sim 10^8$ yr, and it is in a 2:1 MMR with both resonance angles $\theta_1=\lambda_1-2\lambda_2+\varpi_1$ and $ \theta_2=\lambda_1-2\lambda_2+\varpi_2$ librating around $0^{\circ}$.  The evolution of semimajor axes $a_1$ and $a_2$, eccentricities $e_1$ and $e_2$ and resonance angles $\theta_1$ and $\theta_2$ are shown in the right panel of Fig. \ref{fig:RV2190}. Notice that although $e_2$ is relatively small in the first observed epoch,  it fluctuates because of the large libration amplitudes of the resonance angles. The maximum of $e_2$ is about 0.23 and the minimum is about 0.07. 

\begin{figure}
%\epsscale{1.0}
%\plotone{f1.eps}
\centering
\includegraphics[width=0.95\textwidth]{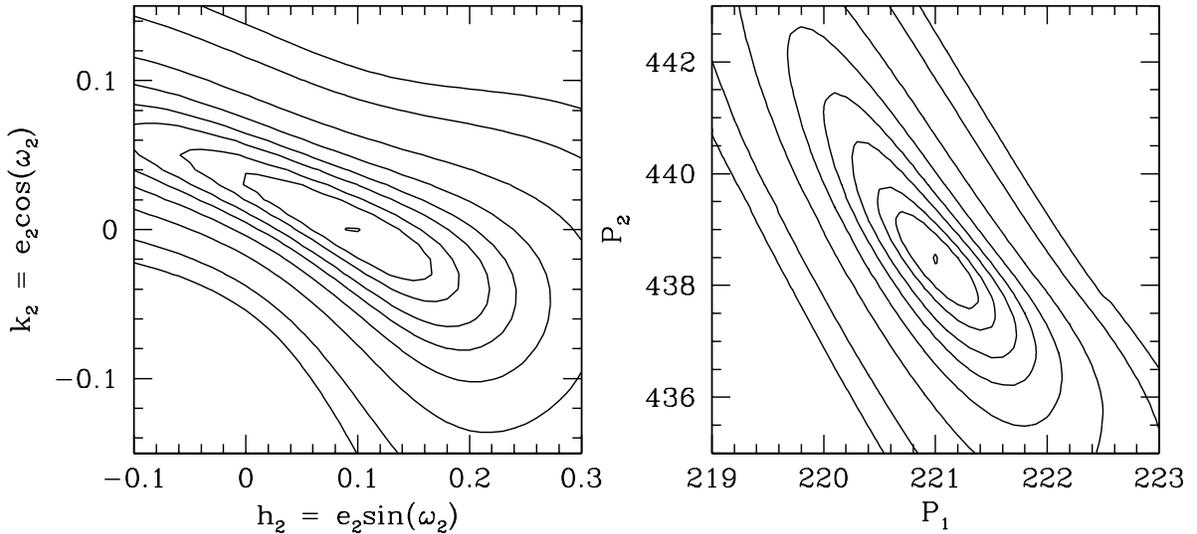}
\caption{ $\chi^2_{\nu}$ contours (1.165, 1.18, 1.195, 1.22,  1.25, 1.3, 1.4, 1.6, 1.8) in the $h_2-k_2$ and $P_1-P_2$ grids for 2:1 MMR coplanar edge-on fits. The $\chi^2_\nu$ contours in both grids converge smoothly to  a single minimum, suggesting the located minimum is the correct one.
\label{fig:chi9021}
}
\end{figure}

Parameter grid search is necessary to examine whether there are any other minima. The errors given by the covariance matrix $\sqrt{C_{ll}}$ in the LM algorithm suggest that $e_2$, $\omega_2$ and $\mathcal{M}_2$ are the least certain parameters. The mean anomaly $\mathcal{M}_2$ is a time-dependent parameter and it does not directly correlate to the orbital spatial configuration, so it is normally not used for grid searching. We search for local best fits with $h_2=e_2\sin \omega_2$ and $k_2= e_2 \cos \omega_2$ fixed in the $h_2-k_2$ grid, because when $e_2$ becomes sufficiently small,  $h_2$ and $k_2 $ are more natural for representing the orbital configuration. After some experiments we decided the range of $h_2$ and $k_2$ to be ($-$0.1 - 0.3) and ($-$0.15 - 0.15) respectively. The left panel of Fig. \ref{fig:chi9021} shows the $\chi^2_{\nu}$ contours (1.165 - 1.80) for the edge-on fits in $h_2-k_2$ grid. The  best fit  is the only minimum in $\chi^2_\nu$ , with $\chi^2_\nu$ = 1.16 for 10 adjustable parameters. The smooth contours show that the $\chi^2$ in the parameter space  smoothly converge to the minimum. We also search fits in other grids. The $P_1-P_2$ grid is the most relevant grid to MMR configuration, and it is shown in the right panel of Fig. \ref{fig:chi9021}. Similar to the $h_2-k_2$ grid, there is only one minimum in the parameter space.

\begin{figure}
%\epsscale{1.0}
%\plotone{f1.eps}
\centering
\includegraphics[width=0.95\textwidth]{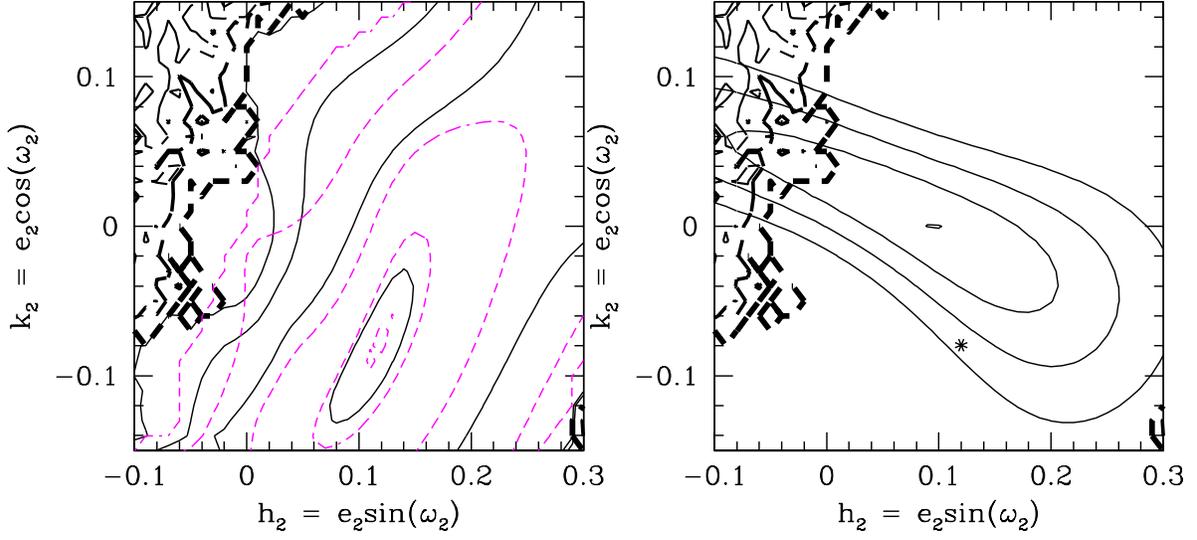}
\caption{  Dynamical stability, libration amplitudes of $\theta_1,\theta_2$ and $\chi_\nu^2$ contours for the 2:1 MMR coplanar edge-on dynamical fits in the $h_2-k_2$ grid. The dashed thick lines in both panels represent the dynamical stability boundary (the thickest one represents 50,000 years and the thiner ones represent shorter survival times). In the left panel, the thin, black, solid curves are contours of $20^{\circ}, 40^{\circ}, 60^{\circ}, 80^{\circ}$ for libration amplitudes of $\theta_1$ with the smaller value closer to center of contours, and the thin, dashed, gray (magenta in the color version)  lines are contours of $20^{\circ}, 40^{\circ}, 60^{\circ}, 80^{\circ}, 179^{\circ}$ for libration amplitudes of  $\theta_2$. The best-fit model is far away from the stability boundary, with libration amplitudes $\Delta\theta_1 \sim 40^{\circ}$ and $\Delta\theta_2 \sim 67^{\circ}$. All stable fits are in 2:1 MMR.
 In the right panel, the black thin lines represent contours of $\Delta\chi_\nu^2 = 0.043,~ 0.114,~0.219$ larger than the minimum (indicated by a dot in the middle of $\Delta\chi_\nu^2$ contours), which are the $1,2,3\sigma$ confidence levels. A large fraction of the $1,2, 3-\sigma$ confidence regions are stable. The star dot is where the fit with the smallest libration amplitudes of resonance angles $\theta_1,~\theta_2$ is located, and it is far away from the best fit. 
\label{fig:h2k290}
}
\end{figure}

\begin{figure}
%\epsscale{1.0}
%\plotone{f1.eps}
\centering
\includegraphics[width=0.95\textwidth]{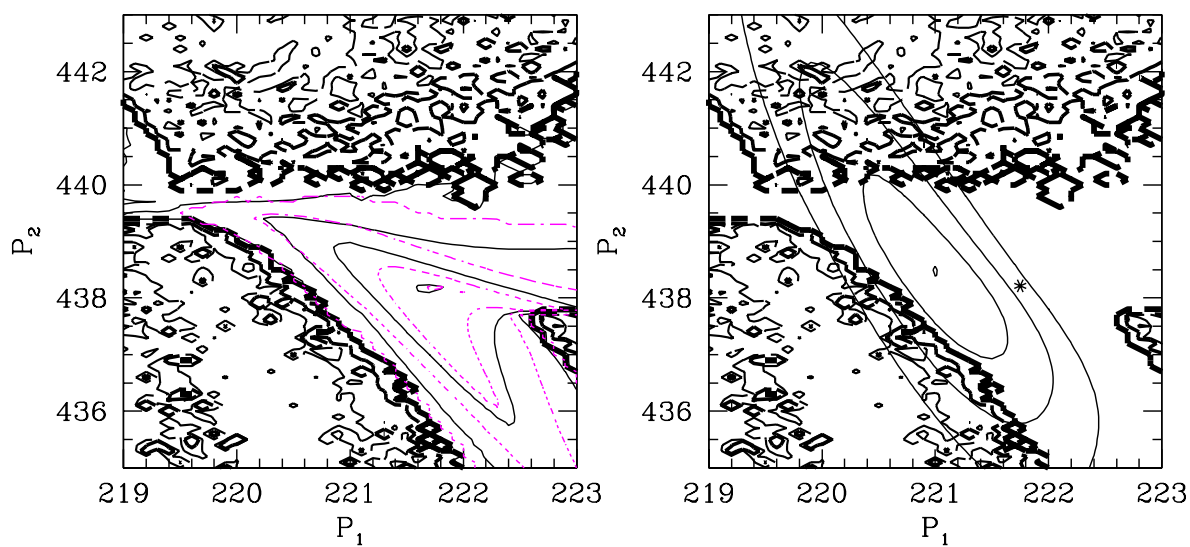}
\caption{   Same as Fig. \ref{fig:h2k290} but for the $P_1-P_2$ parameter grid. In the left panel, the thin, solid, black contour values are $24^{\circ}, 40^{\circ}, 60^{\circ}, 100^{\circ}$, and the thin, dashed, gray (magenta in the color version) contour values are $36^{\circ}, 55^{\circ}, 80^{\circ}, 179^{\circ}$. A large fraction of the $2,3-\sigma$ confidence regions can be excluded due to their unstable nature, but almost all of the $1-\sigma$ region is stable. The fits with the smallest libration amplitudes is  far away from the best fit.
\label{fig:p1p290}
}
\end{figure}

We use the symplectic integrator SyMBA \citep{Duncan1998} to perform a
direct $N$-body integration with a maximum time of 50,000 yr  for each
local best-fit model in the grids. The initial semimajor axes of the
inner and outer planet are about 0.75 AU and 1.18 AU,
respectively. For the system to be considered stable, two criteria
have been set: (1) the maximum distance of a planet to the star must
be less than 3 AU (5 AU for 3-planet models in \S 4.2) and the minimum
distance must be larger than 0.075 AU; (2) the distance between the
planets must be larger than the sum of their physical size assuming
their mean density of 1 g $\rm{cm}^{-3}$. The minimum and maximum
distances from the star adopted in criterion (1) are sufficiently
different from the initial semimajor axes of the planets that the
system is clearly unstable if these limits are exceeded. The dynamical
properties of the fits in the $h_2-k_2$ grid are shown in
Fig. \ref{fig:h2k290}. In the left panel, the thick dashed lines are
contours of survival time of the integrations, with the  thickest line
corresponding to 50,000 yr (the maximum integration time) and the
thinnest line corresponding to 2,000 yrs. The blank region is the
stable region in which all fits are stable. All stable fits are in 2:1
MMR with $\theta_1$  librating around $0^{\circ}$. Most stable fits
have $\theta_2$ librating around $0^{\circ}$ but some fits close to
the dynamical stability boundary have $\theta_2$ circulating. The thin
solid black and thin dashed gray (magenta in the color version) curves
represent the contours of libration amplitudes for $\theta_1$ and
$\theta_2$, respectively. In the right panel, the thick dashed lines
are the stability boundary, and the thin curves are $\chi^2_{\nu}$
contours corresponding to the $1\sigma$, $2\sigma$ and $3\sigma$
contours of confidence levels based on $\Delta\chi^2$ of 2.3, 6.17,
11.8 (or $\Delta\chi^2_{\nu}$ of 0.043, 0.114, 0.219) larger than the
minimum for 2 parameters \citep{Press1992}. The star point represents
the fit with the smallest libration amplitudes of both resonant angles
($\Delta\theta_1\approx14^{\circ}, \Delta\theta_2\approx16^{\circ}$),
and it is outside the $2\sigma$ confidence region. The best-fit model
is far away from the stability boundary, with libration amplitudes
about $40^{\circ}$ and $67^{\circ}$ for $\theta_1$ and $\theta_2$,
respectively (see also Fig. \ref{fig:RV2190}). A large fraction of the
$1,2,3\sigma$ confidence regions  are in the stable region, and a
small fraction of fits in $1-\sigma$ confidence region are
unstable. We do similar analysis in the $P_1-P_2$ grid, and the
results are shown in Fig. \ref{fig:p1p290}, with the same contents as
the $h_2-k_2$ grid. The $2,~3\sigma$ confidence regions are narrowed
by the stability boundaries compared to $h_2-k_2$ grid, but almost all
fits in the $1-\sigma$ confidence region are stable.

\subsubsection{Coplanar Inclined  Fits}
\begin{figure}
\centering 
%\epsscale{0.75}
\includegraphics[width=0.75\textwidth]{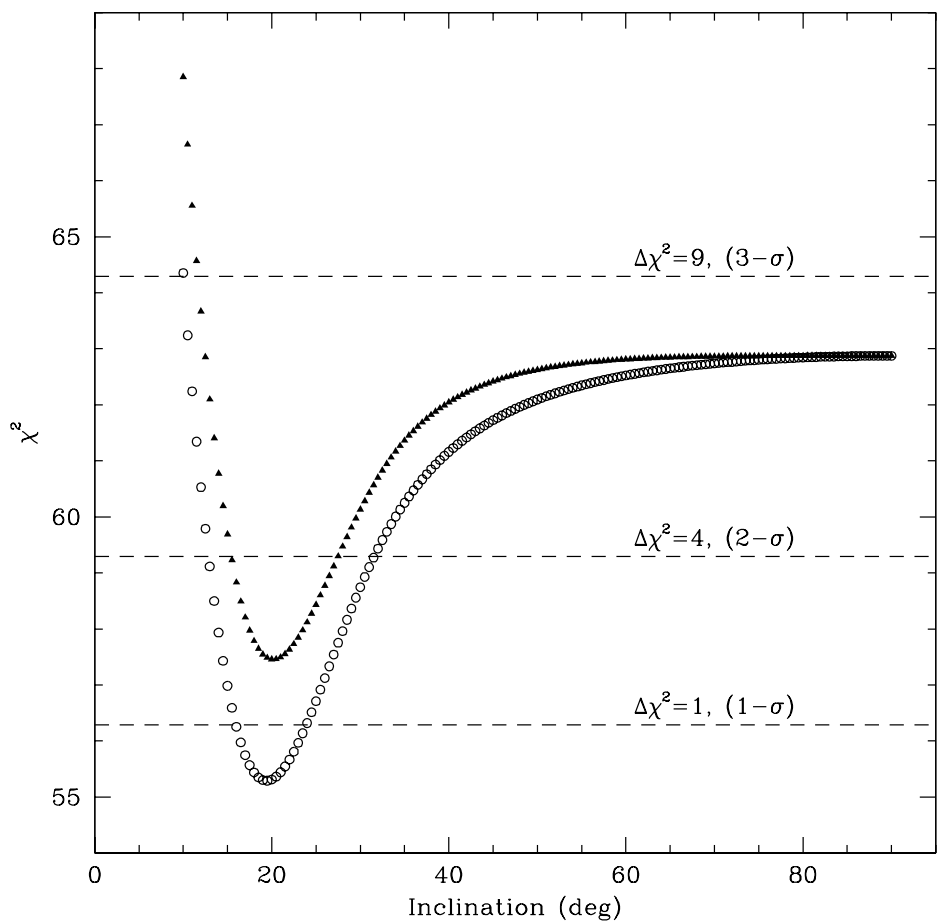}
%\plotone{./fig/chisq_inc21.pdf}
\caption{ $\chi^2$  of the coplanar 2:1 MMR best fit as a function of inclination. The \emph{open circles} represent $\chi^2$  and the \emph{filled triangles} represent ($\chi^2-2\ln \sin i$) as a function of inclination for 2:1 MMR coplanar  best fits.  The dashed lines represent  $\Delta\chi^2$ values of 1.0, 4.0, 9.0 larger than the minimum $\chi^2$ value around $20^{\circ}$, which are the $1,2,3-\sigma$ confidence levels. The minimum at about $20^{\circ}$ is statistically significant, indicating that an inclined solution is preferred. 
\label{fig:chisqinc}
}
\end{figure}

\begin{figure}
\centering 
\includegraphics[width=0.75\textwidth]{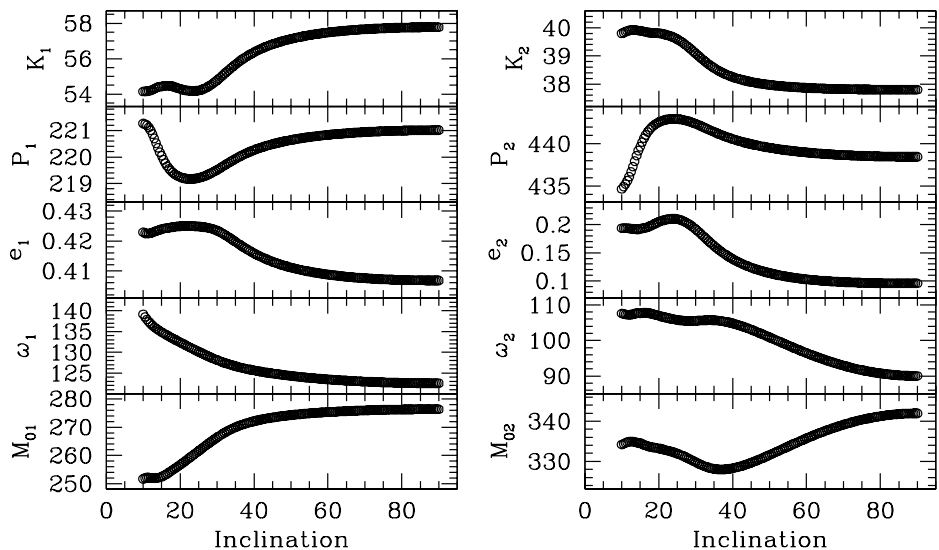}
\caption{ Fitting parameters of planet 1 and planet 2  as a function of inclination for coplanar 2:1 MMR best fits. The inflection points near $20^{\circ}$ inclination of some fitting parameters, e.g., $P_1,P_2$ and $e_2$, may be the cause for the discontinuities of $\chi^2_{\nu}$ contours in $h_2-k_2$ and $P_1-P_2$ grid for coplanar inclined fits that allow inclination to vary (Fig. \ref{fig:chiifree21}).
\label{fig:parainc}
}
\end{figure}

Dynamical fitting can be sensitive to the true masses of planets, so it may be possible to determine the inclinations of the planetary orbits. In this subsection, we assume that the planets are on coplanar orbits. We allow the system's orbital inclination relative to the sky plane to vary together with the other orbital parameters. $\chi^2$ and fitting parameters of the best fit as a function of inclination are shown in Fig. \ref{fig:chisqinc} and Fig. \ref{fig:parainc}. In Fig. \ref{fig:chisqinc}, the open circles are $\chi^2$ values, and the dashed lines represent 1, 2 and 3 $\sigma$ confidence levels, which are $\Delta\chi^2$ of 1.0, 4.0 and 9.0 larger than the minimum $\chi^2$. The $\chi^2$ changes slowly until the inclination drops below about $40^{\circ}$. The decreasing $\chi^2$ stops at about $20^{\circ}$, then increases  with  decreasing inclination. The minimum of $\chi^2$ at about $20^{\circ}$ inclination is $\Delta \chi^2 \approx 8$ lower than the value of $\chi^2$ for the edge-on best-fit model, which is a close to $3\sigma$ improvement over the edge-on best fit. The parameters of the best fit also change with the inclination as shown in Fig. \ref{fig:parainc}, and there are  inflection points  near $20^{\circ}$ inclination for some fitting parameters. In Bayesian inference, an isotropic distribution of orbit normals implies a prior of $\sin i$ for the inclination and an effective likelihood function $\mathcal{L}_e = \sin i \times \mathcal{L}$, where $\mathcal{L} \propto \exp(-\chi^2/2)$. The equivalent quantity ($\chi^2-2\ln\sin i$) is shown as a function of inclination by the filled triangles in the upper panel of Fig. \ref{fig:chisqinc}.  Although the prior reduces the effective likelihood when the orbit is highly inclined from the line of sight, the effective likelihood function also reaches a maximum at about $20^{\circ}$ inclination, similar to the original likelihood function. 

\begin{figure}
\centering 
\includegraphics[width=0.95\textwidth]{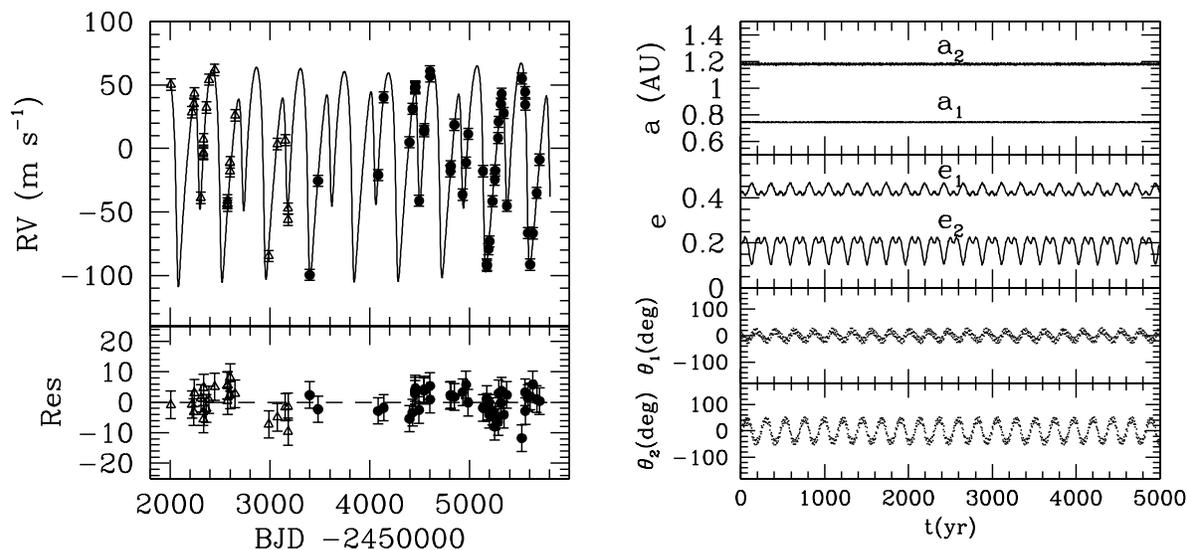}
\caption{ RV curve and residuals of the 2:1 MMR coplanar $20^{\circ}$ inclined best fit and its dynamical evolution. For the inclined dynamical best fit, the most significant differences compared to the edge-on fit are the larger eccentricity of the outer planet at the first epoch, and that the masses of the two planets become almost equal.
\label{fig:RV2120}
}
\end{figure}

The best fit is at near $20^{\circ}$ inclination, and its fitting parameters  are listed in the third part of Table \ref{tab:best21}.  The $\chi^2_{\nu}$ of 1.08 is close to 1, and the rms of $4.09\mps$ is consistent with the estimated stellar jitter $(4.2\mps)$. The masses of the inner and outer planets are 4.78 and 4.80 $M_{\rm{J}}$ respectively, where $M_{\rm{J}}$ is the mass of Jupiter. The RV curve is shown in the left panel of Fig. \ref{fig:RV2120}.  The best fit is dynamically stable for at least $10^8$ yr and is in the 2:1 MMR with both resonance angles, $\theta_1$ and $\theta_2$,  librating around $0^{\circ}$, as shown in the right panel of Fig.  \ref{fig:RV2120}. 
The major difference from the edge-on best fit is that the eccentricity of the outer planet is larger and the masses of the two planets are almost equal. As shown in the right panel of Fig. \ref{fig:RV2120},  the librating behavior of $e_1$ and $e_2$ is similar to that of the edge-on best fit, but the libration period is shorter and the average value of $e_1$ is slightly larger than those of the edge-on best fit.

\begin{figure}
\centering
\includegraphics[width=0.95\textwidth]{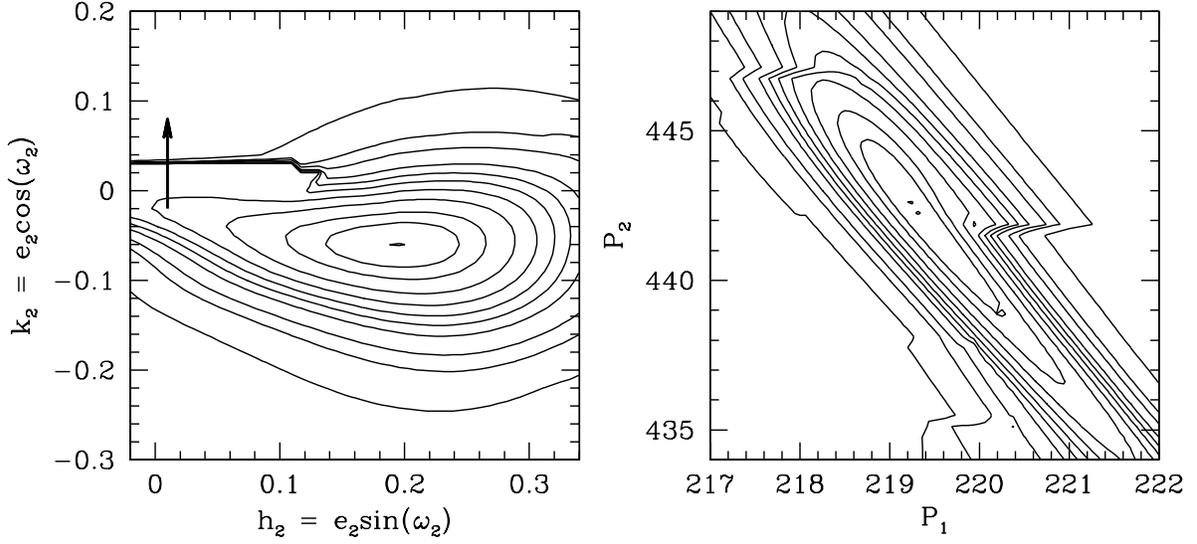}
\caption{  $\chi^2_{\nu}$ contours (1.044, 1.065, 1.09, 1.13, 1.17, 1.21, 1.25, 1.3, 1.4, 1.5, 1.8) in the $h_2-k_2$ and $P_1-P_2$ grids for coplanar inclined 2:1 MMR fits allowing $i$ to float. As in Fig. \ref{fig:chi9021} (edge-on case) we see that there is only a single minimum for each grid. 
\label{fig:chiifree21}
}
\end{figure}

\begin{figure}
%\epsscale{1.0}
%\plotone{f1.eps}
\centering
\includegraphics[width=0.95\textwidth]{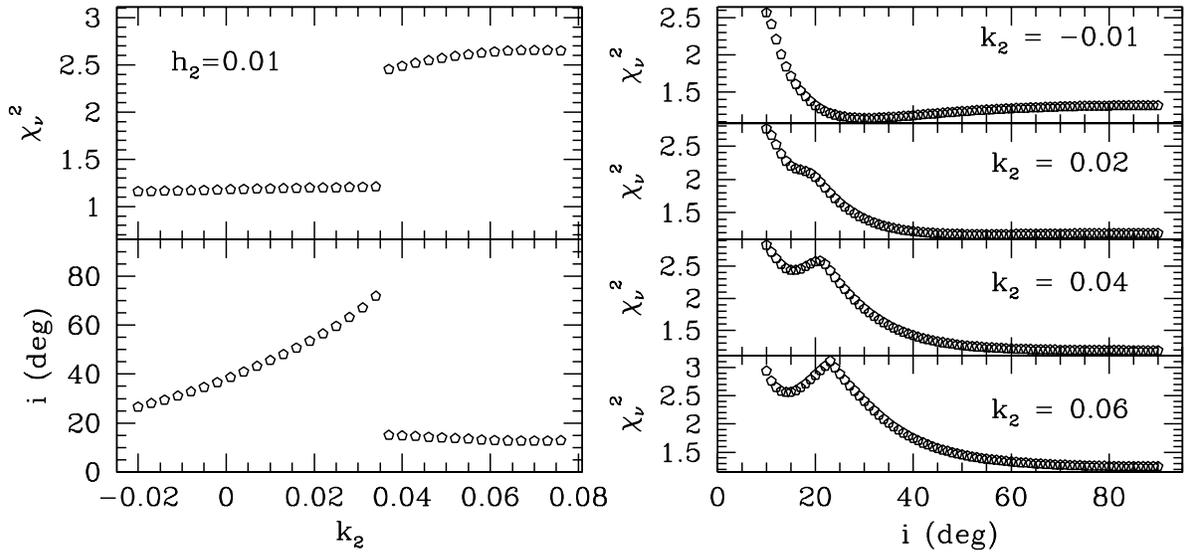}
\caption{ Left: Changes in inclination and $\chi^2_{\nu}$ for the fits
along the arrow marked in the left panel of
Fig. \ref{fig:chiifree21}.  Right: Illustrations that $\chi^2_{\nu}$
as a function of inclination shows a second minimum around
$i\approx15^{\circ}$ when $k_2 \ga 0.025$ for $(h_2,k_2)$ along the
arrow in the left panel of Fig. \ref{fig:chiifree21}.
\label{fig:cross}
}
\end{figure}

Similar to the edge-on fits in \S~4.1.1, we conduct a grid search
around the coplanar $20^{\circ}$ inclined best fit allowing the
inclination to {\it vary}.
Fig. \ref{fig:chiifree21} shows the $\chi^2_{\nu}$ contours for the
$h_2-k_2$ and $P_1-P_2$ grids.
Similar to edge-on fits, only one minimum is found for each grid, and
the inclinations $i$ of the fits near the minimum are not far from
$20^{\circ}$.
However, the contours show discontinuities, which are not present in
the edge-on fits.
% 
%%%%%%    explain the discontinuities   %%%%%%%%%%%%%%
The discontinuities in $\chi^2_{\nu}$ and the free fitting parameters
are illustrated in the left panel of Fig. \ref{fig:cross} by
extracting $\chi^2_{\nu}$ and inclination (one of the free fitting
parameters) along the arrow (from $[h_2=0.01,k_2=-0.02]$ to
$[h_2=0.01, k_2=0.08]$) in the left panel of Fig. \ref{fig:chiifree21}.
The reason for the discontinuities is the appearance of a second local
$\chi^2$ minimum with respect to the free fitting parameters when
$k_2 \ga 0.025$.
The right panel of Fig. \ref{fig:cross} shows $\chi^2_{\nu}$ as a
function of inclination with $h_2 = 0.01$ and different fixed $k_2$
values taken along the arrow in the left panel of
Fig. \ref{fig:chiifree21}.
For $k_2 \la 0.025$, there is a single minimum with $\chi_\nu^2 \sim
1.15$ at an inclination that increases with increasing $k_2$.
For $k_2 \ga 0.025$, a second minimum with $\chi_\nu^2 \ga 2.3$
appears around $i\approx15^{\circ}$.
Because the starting condition of $\chi^2$ minimization is small
inclination along the arrow, the fit is trapped in the minimum around
$i \approx 15^{\circ}$ when $k_2 \ga 0.035$.
The discontinuities in the $P_1-P_2$ grid shown in the right panel of
Fig. \ref{fig:chiifree21} can be explained similarly.
% %%%%%%%%%%%

\begin{figure}
\centering
\includegraphics[width=0.95\textwidth]{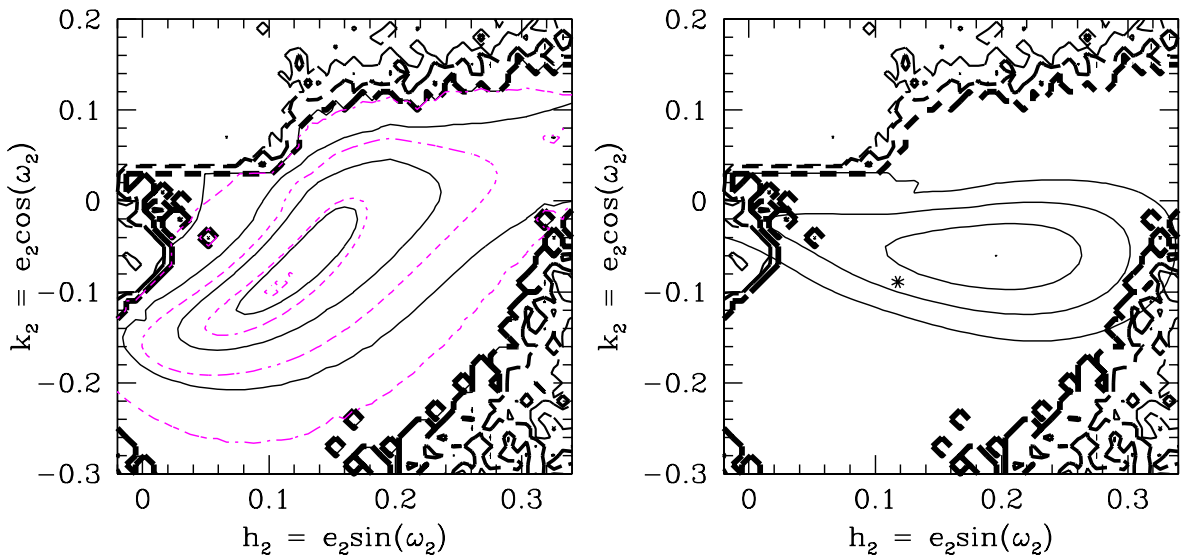}
\caption{  Dynamical stabilities and properties of 2:1 MMR coplanar inclined fits for which the inclination, $i$, is allowed to float in the $h_2-k_2$ parameter grid. The donations of lines are the same as Fig. \ref{fig:h2k290}. In the left panel, the thin, solid, black contour values are $20^{\circ}, 30^{\circ}, 40^{\circ}, 60^{\circ}$, and the thin, dashed, gray (magenta in the color version) contour values are $20^{\circ}, 40^{\circ}, 60^{\circ}, 80^{\circ}$. As in Fig. \ref{fig:h2k290} for the edge-on case, the best-fit model is far away from the stability boundary. However, for this inclined case, an even larger fraction of the $1,2,3-\sigma$ confidence regions are stable.
\label{fig:h2k2ifree}
}
\end{figure}

\begin{figure}
%\epsscale{1.0}
%\plotone{f1.eps}
\centering
\includegraphics[width=0.95\textwidth]{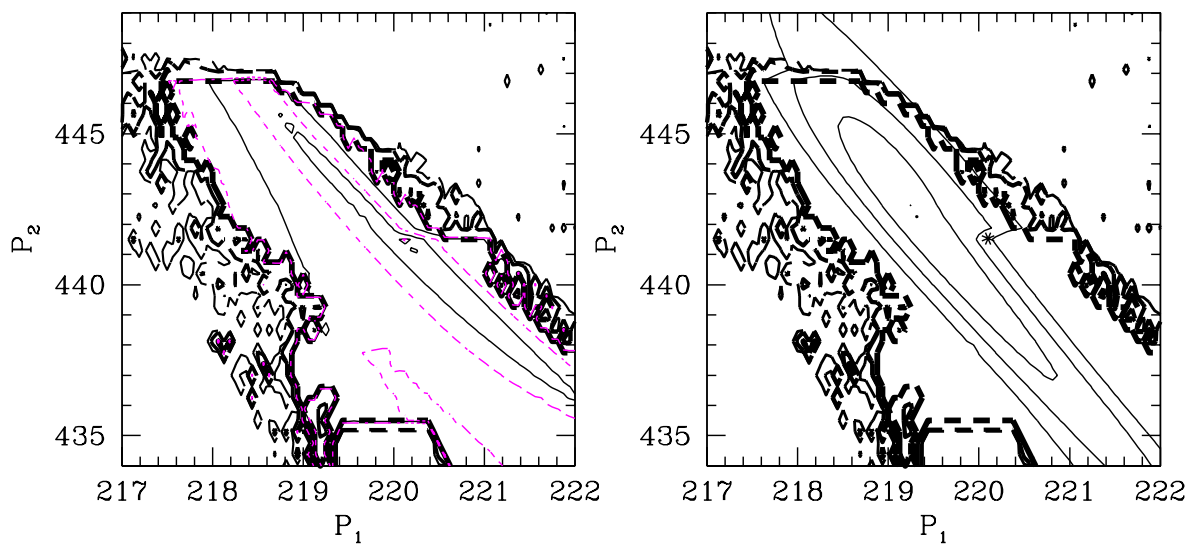}
\caption{  Same as Fig. \ref{fig:h2k2ifree} but for the  $P_1-P_2$  grid. In the left panel, the thin, solid, black contour values are $13^{\circ}, 20^{\circ}, 40^{\circ}$, and the thin, dashed, gray (magenta in the color version) contour values are $22^{\circ}, 50^{\circ}, 80^{\circ}$. Again, we see that these inclined solutions have a greater fraction of the $1,2,3-\sigma$ confidence regions are stable. In addition, it should be noted that the libration amplitudes are smaller in terms of the stable region (compare to the edge-on case), presumably because these more massive planets more easily become unstable at high libration amplitudes.
\label{fig:p1p2ifree}
}
\end{figure}

Fig. \ref{fig:h2k2ifree} shows the dynamical properties of fits in the $h_2-k_2$ grid, with similar contents as Fig. \ref{fig:h2k290} of edge-on fits. Similar to the edge-on case, the stable region is the region where there are libration amplitude contours (the thin solid black and thin dashed gray curves). The $\chi^2$ minimum is also far from the dynamical stability boundary, with libration amplitudes about $30^{\circ}$ and $52^{\circ}$ for $\theta_1$ and $\theta_2$, respectively (see also Fig. \ref{fig:RV2120}). Most fits in the $3\sigma$ confidence region and almost all fits in the $2\sigma$ confidence region are in the stable region, different from the edge-on fits where there is still a small fraction of fits in the $1, 2,3\sigma$ confidence region being unstable. All fits in the stable region are in 2:1 MMR with both $\theta_1$ and $\theta_2$ librating around $0^{\circ}$, and the fit with the smallest libration amplitudes of both resonant angles ($\Delta\theta_1 \approx 14^{\circ}, \Delta\theta_2\approx16^{\circ}$) is near the $2\sigma$ confidence level. Fig. \ref{fig:p1p2ifree} shows the dynamical properties of fits in the  $P_1-P_2$ grid. The dynamical properties and the $\chi^2$ statistics of  this grid are  similar  to those of  the $h_2-k_2$ grid.

\subsubsection{Mutually Inclined Fits}

Very few extrasolar planetary systems have successfully had their mutual inclinations measured by radial velocity alone. The most familiar case would be the GJ 876 planetary system, but even in this case the conclusions are not yet consistent. Based on the combination of RV and astrometry data, \citet{Bean2009} showed that the mutual inclination between   the  planets GJ 876 b and c is ${5.0^{\circ}}^{+3.9^{\circ}}_{-2.3^{\circ}}$. Based on pure RV data, \citet{Correia2010} showed that the mutual inclination is less than $2^{\circ}$, and an updated analysis  by \citet{Baluev2011} limited it by $5^{\circ}-15^{\circ}$. 

\begin{figure}
\centering
\includegraphics[width=0.95\textwidth]{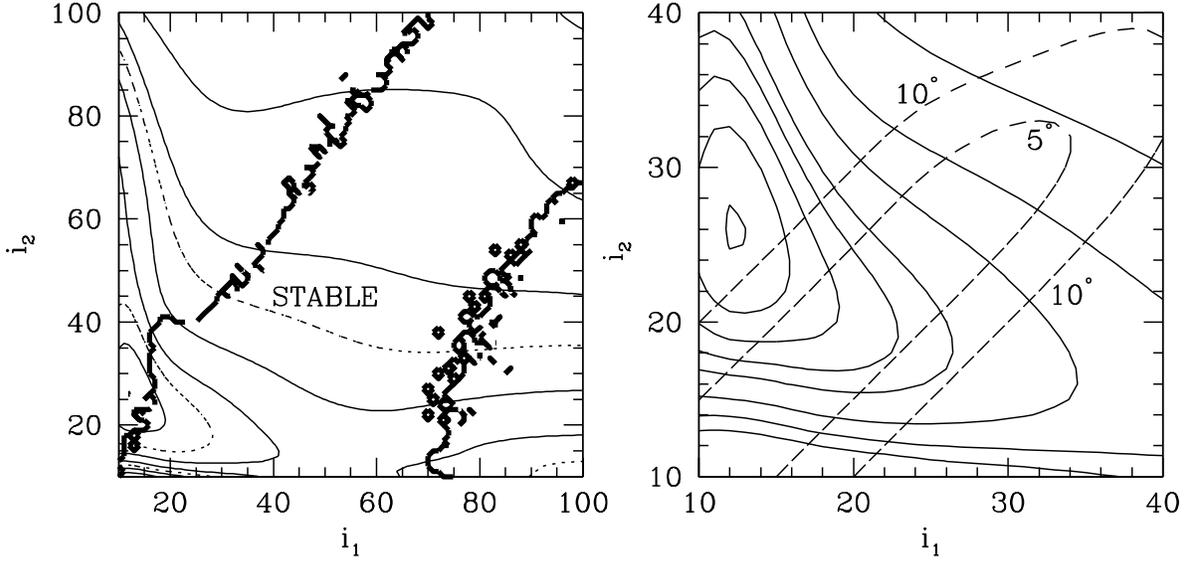}
\caption{ $\chi^2_{\nu}$ contours and dynamical stability test in the $i_1-i_2$ grid for mutually inclined 2:1 MMR fits. The left panel shows the dynamical stability boundary (thick dashed line) and $\chi^2_{\nu}$ contours (thin curves with value: 1.06, 1.08, 1.105, 1.13, 1.16, 1.181, 1.2 1.23), with the thin dot lines being the $1\sigma$ and $2\sigma$ confidence levels (note that the minimum model indicated by a dot is in the lower left corner of the grid, and is just outside the stable boundary). The right panel is the expansion of the lower left part of the grid, in which curves represent $\chi^2_{\nu}$ contours and dashed lines represent the $\Delta i$ contours of $5^{\circ}$ and $10^{\circ}$, respectively. The distribution of the $\chi^2_\nu$ contours against the $\Delta i$ contours is such that we {\it cannot} constrain the mutual inclination between the planets.
\label{fig:i1i221}
}
\end{figure}

Since we are able to constrain the inclination of the HD 82943 system if we assume coplanar orbits, it is interesting to try to constrain the mutual inclination between the two planets in the system.   We use the best-fit coplanar model of 2:1 MMR as an initial guess. After convergence of the LM method, $\chi^2_{\nu}$ reached 1.10 and the inclinations of the planets relative to the sky plane are about $i_1=12.3^{\circ}$ and $i_2=26.0^{\circ}$, respectively. The $\chi^2$ becomes better, but the $\chi^2_{\nu}$ becomes worse after 2 more fitting parameters are introduced. Moreover, the fit is unstable on the order of 10,000 yr. A systematic parameter grid search show that we cannot constrain the mutual inclination. Fig. \ref{fig:i1i221} shows the results of the $i_1-i_2$ grid search coupled with  dynamical analysis. In the left panel, the thin curves are the $\chi^2_{\nu}$ contours with the thin dotted lines representing  the minimum (near the lower left corner of the grid) and the 1 and $2\sigma$ confidence levels. The thick dashed line is the dynamical stability boundary, and the region in the middle of the grid is the stable region. The $\chi^2$ minimum located in the lower left corner of the grid is dynamically unstable and is shallow  in parameter space. The right panel is an expansion of the lower left part of the grid.  The solid curves represent $\chi^2_\nu$ contours and the thin dashed lines  represent the contours of mutual inclination of  $10^{\circ}$ and $5^{\circ}$. Note that the mutual inclination is determined by $i_1,~i_2$ and the mutual longitude of ascending node $\Delta\Omega$. The $\chi^2$ contours cross the  region within $10^{\circ}$ and $5^{\circ}$ mutual inclination, showing that fits of different mutual inclination have the same quality and thus cannot be constrained by $\chi^2$ statistics. The stability test cannot constrain the mutual inclination either. In the $e_2-\omega_2$ grid searching, another local minimum was found with $\chi^2_{\nu}$ of 0.96. However this orbital configuration is extremely unstable because the system is highly mutually inclined with one of the ``planets''  having the mass of a brown dwarf. Based on the current data,  the coplanar inclined best fit is already adequate, and  it seems difficult to find better fits in mutually inclined configuration of 2:1 MMR.

\subsubsection{Summary for 2:1 Resonance Fits}
In summary, the coplanar inclined best fit results in significant improvement in $\chi^2_{\nu}$ statistics ($\chi^2_{\nu}=1.08$ of the $20^{\circ}$ inclined best fit compared to $\chi^2_{\nu}=1.21$ of the edge-on best fit and $\chi^2_{\nu}=1.38$ of Keplerian best fit).  Almost all fits inside the $2\sigma$ confidence region are stable, and all stable fits have both resonant angles $\theta_1$ and $\theta_2$ librating around $0^{\circ}$.  The stable 2:1 MMR configuration is robust for the HD 82943 system because all good fits are stable and in the 2:1 MMR. The fits with the smallest libration amplitudes of both resonant angles ($\Delta\theta_1 \approx 14^{\circ}, \Delta\theta_2\approx16^{\circ}$) are about $2\sigma$ from the best fit (with $\Delta\theta_1 \approx 30^{\circ}, \Delta\theta_2\approx52^{\circ}$) in the parameter grids, suggesting that the system does not favor small-libration-amplitudes configuration. On the other hand, we cannot solve for the {\it mutual }inclination for the HD 82943 system.

\subsection{3-Planet Fits}
\label{3planet}

The Laplace resonance configuration is well known as the double 2:1 MMR among the Galilean satellites Io, Europa, and Ganymede. In extrasolar planetary systems, the Laplace resonance may also play an important role in various MMR configurations. For example, the GJ 876 system \citep{Rivera2010} and the HR8799 system \citep{Marois2010} are suspected to contain planets in Laplace resonance. The existence of a third outer planet in a Laplace resonance with the two existing planets for the HD 82943 system was suggested by \citet{Beauge2008}. It is interesting to examine the viability of the 3-planet (Laplace resonance) fits with the new Keck data. 

\begin{figure}
\centering 
\includegraphics[width=1.0\textwidth]{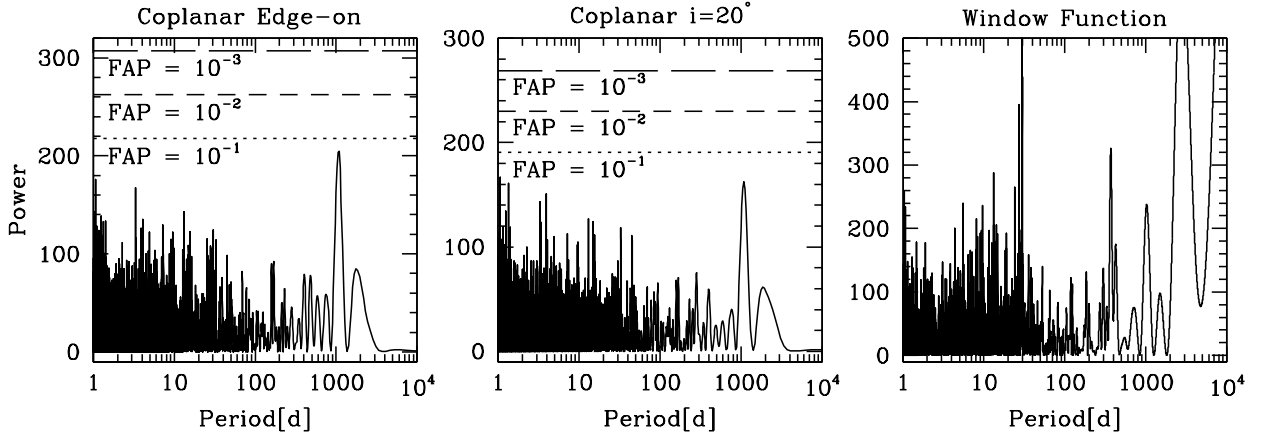}
\caption{ Periodograms of residuals from best fits on 2:1 MMR configuration (the left panel is for residuals from edge-on best fit, and the middle panel is for residuals from $20^{\circ}$ inclined best fit). The right panel shows spectrum of the window function. For both edge-on and $20^{\circ}$ inclined case, the peaks at $\sim 1,100$ days are of low amplitudes, hence do not pass the analytical false-alarm test. Moreover, the window function has significant amplitude at about 1,100 days, indicating that peaks around 1100 days in periodograms result from  structured systematic noise. Finally, the analytic FAPs are verified by a separate false alarm analysis using a complementary bootstrapping approach. As such, there is little convincing evidence for the presence of a third planet.
\label{fig:periodogram}
}
\end{figure}

First, we input the residuals from the 2:1 MMR best fits of coplanar edge-on and $20^{\circ}$ inclined orbits to the Lomb-Scargle periodogram function (e.g., \citealp{Press1992}) implemented in the Systemic Console \citep{Console}. The power spectra as a function of period are shown as the left and middle panels of Fig. \ref{fig:periodogram}. There is a peak at around 1100 days for both power spectra of edge-on and $20^{\circ}$ inclined orbits. However, both peaks have false alarm probability (FAP) larger than 10\%, i.e., the peak is lower than the 10\% FAP line which is the lowest dashed lines in the figure. Conventionally, the statement of having a new planet should be based on having a periodogram power spectrum peak at least higher than the line corresponding to $\rm{FAP}=10^{-2}$ (e.g.,~\citealp{Marcy2005b}), which is the middle dashed line in the figure. 
The right panel of Fig. \ref{fig:periodogram} shows the power spectrum of the window function which evaluates the periodicity contributed to the data from the choice of observational epochs. There is a peak at about 1100 days, which corresponds to  an observational period of about 3 years. The coincidence for the three analysis in Fig. \ref{fig:periodogram} all showing peaks at about 1100 days  hints that the periodic signals in the residuals are partially due to the structured systematic noise.\footnote{
The periodogram of the window function in Figure \ref{fig:periodogram}
also shows peaks at about $1$ month and $1$ year, but there are no
corresponding peaks in the periodogram of the residuals, because the
periodogram of the residuals depends on the properties of the
residuals as well as the window function. The periodogram of the
window function is best used as a guide to those frequencies that
should be treated with caution, not as a prediction of the locations
of all spurious peaks.
}
To verify the validity of the analytic FAPs calculated by Systemic, we have conducted a separate false alarm analysis using a complementary bootstrapping approach, similar to that of \citet{Wright2007} and \citet{Marcy2005b}.  
We randomly redrew the velocity residuals to the edge-on and $20^{\circ}$ inclined 2:1 resonant cases (with replacement), maintaining the temporal spacing of the observations, 1000 times.  In each case we calculated the height of the tallest peak in the periodogram and compared to the tallest peak in the periodogram of the unscrambled data.  In our unrestricted analysis (periods from 1 day to 10,000 days), we find that peaks near 1 day and 1100 days are tall, which result from the window function of this data set. Our FAP for the peak around 1100 days is 3.3\% for edge-on case and 8.5\% for $20^{\circ}$ inclined case,  which are lower than the analytic FAP but still not comparable to $10^{-2}$.
We then restrict our  analysis to periods between 2 and 900 days, to avoid the tall peaks near 1 day and 1100 days.
We find in the $20^{\circ}$ inclined case, 26\% of our synthetic data sets had a peak taller than the tallest seen in the actual data set, indicating a FAP value $\sim 26\%$.  In the edge-on case, we find FAP $\sim 20\%$.  
Thus, there is no evidence for the existence of a third planet in the Keck data.

\begin{figure}
\centering 
\includegraphics[width=0.95\textwidth]{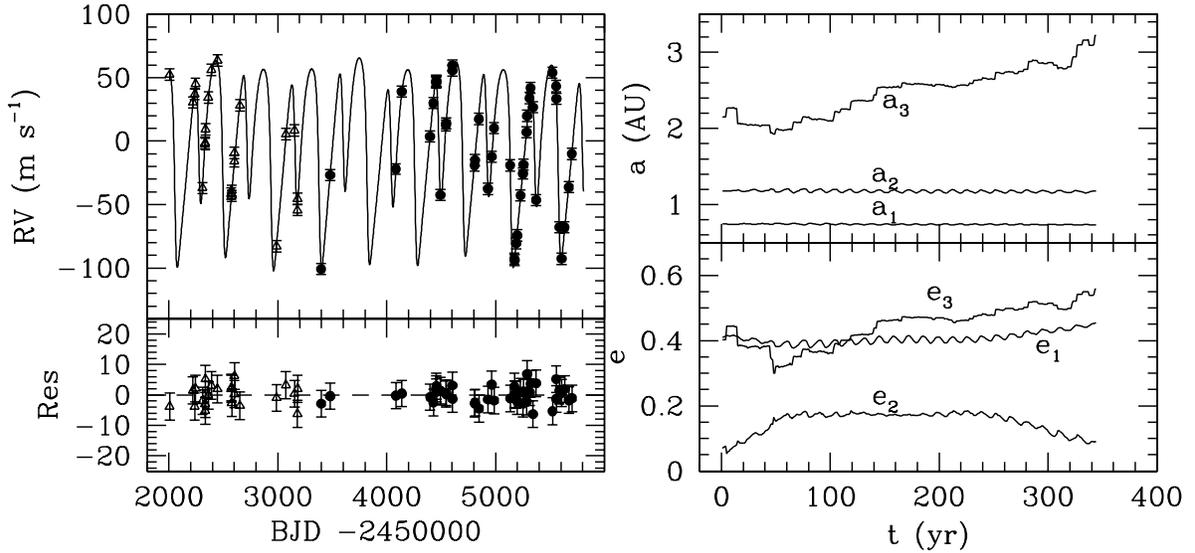}
\caption{ RV curve and residuals of 3-planet best fit and its dynamical evolution. This best-fit 3-planet model becomes unstable within a few hundred years, primarily due to the high eccentricity of the third planet driving strong interactions with the inner planets.
\label{fig:3p}
}
\end{figure}

Nevertheless, we try to fit a 3-planet model and look for any configurations associated with the Laplace resonance. Assuming coplanar edge-on orbits, the residuals from the coplanar edge-on best fit are treated  as a new data set. A Keplerian orbit is fitted to the residuals with a period of about 1100 days. Then we input the initial guess of the 3-planet model from the sum of the edge-on two-planet best fit and the Keplerian fit of the third planet to our dynamical fitting code, allowing  all parameters (except the fixed inclination $i=90^{\circ}$) to vary. A local minimum is found with a $\chi^2_{\nu}$ of 0.59 and rms of $2.91\mps$. The parameters of this local best fit are listed in Table \ref{tab:3p} and its RV curve and residuals are shown in the left panel of Fig. \ref{fig:3p}. The period of the third planet is 1077 days, close to the peak in the periodogram, and $e_3$ is  0.402. This fit is dynamically unstable in a few hundred  years as shown in the right panel of Fig. \ref{fig:3p}. Because of the high eccentricity  $e_3$ of the third planet, its orbit  is easily perturbed by the two massive planets inside and thus becomes unstable.  The large uncertainties of some fitting parameters of the third planet from the covariance matrix $\sqrt{C_{ll}}$ suggest that there may be other minima.  Starting from this local best fit, we search in the $h_3-k_3$ grid, and found two other local minima with slightly higher $\chi^2_{\nu}$. So the fit in Table \ref{tab:3p} is likely the global best fit of the 3-planet coplanar edge-on configuration. Similar to the best fit, the other two local minima are dynamically unstable in a few hundred years. In fact, dynamical stability test in the $e_3-\omega_3$ grid shows that all fits with $e_3 \gtrsim 0.06$ are unstable. In order to find fits associated with Laplace resonance, we search the $P_2-P_3/P_2$ grids starting separately  from the three minima found in the $h_3-k_3$ grid. We do not find local minimum near the nominal Laplace configuration, i.e., $P_3/P_2 \approx 2.0$.  Fits that are close to the nominal Laplace resonance configuration have  $\chi^2_{\nu} \gtrsim 0.84$, which is much larger than that of the 3-planet best fit, and they are dynamically unstable.  Finally, we allow the inclination to float assuming coplanar orbits, and then fit the data starting with the coplanar edge-on best fit. Similar to the coplanar 2:1 MMR fits, the $\chi^2$ shows a minimum ($\chi^2_{\nu}=0.54$) at about $20^{\circ}$. 

In summary for the 3-planet fits,  the periodograms of the residuals
from the  2:1 MMR best fits do not provide  evidence for the existence
of a third planet. Additionally, the $\chi^2_{\nu}$ of the best fit is
significantly lower than 1.0, which together with the goodness of the
2:1 MMR best fit ($\chi^2_{\nu} = 1.08$), hints that the  3-planet
model results in over fitting the current data (i.e., the 3-planet
model has too many parameters and its fit to the data is ``too
good''\footnote{
According to the chi-square test, the probability that $\chi^2_{\nu}$
does not exceed $0.59$ of the best edge-on 3-planet fit is less than
$\sim 1\%$, if our estimates of the uncertainties (internal and
stellar jitter) are correct.
}). Finally, dynamical exploration shows that all good 3-planet best
fits are dynamically unstable, and there is no good fit corresponding
to the  nominal Laplace resonance configuration.

\subsection{1:1 Resonance Fits}

\citet{Laughlin2002}  pointed out  that a 1:1 eccentric resonance configuration could be found in extrasolar planetary systems. The eccentric 1:1 resonant configuration can be generated by initially placing a planet  in a circular orbit and the other planet  in a highly eccentric orbit, with the period ratio nearly 1.0. The system maintains a stable configuration with angular momentum exchange between the two planets (i.e., the eccentricities are oscillating). \citet{Gozdziewski2006} reported a group of fits with 1:1 resonance that fit the combined RV data of CORALIE and  Keck observations of the HD 82943 system  almost as well as the 2:1 MMR fits.

\subsubsection{Coplanar Fits}
First, we explore edge-on coplanar 1:1 resonance fits. We use Keplerian fitting to search for an initial guess. We skip the strongest periodic signal of about 220 days in the original data set and directly fit  with a Keplerian orbit at about 440 days. As we already know that the orbits may have  high eccentricities,  we force the first Keplerian orbit to have a relatively high eccentricity and then check the periodogram of the residuals. A periodicity of about 450 days is identified, and then the second Keplerian orbit with about 450 days is fit.   The Keplerian best fit near the 1:1 resonance is adopted as an initial guess for dynamical fitting. The LM method quickly converges to a fit with $\chi^2_{\nu}$ of  1.69 and rms of $5.14\mps$. The parameters of this local best fit are listed in Table \ref{tab:11} with both $e_1$ and $e_2$  large ($e_1=0.466$ and $e_2=0.654$). Similar to previous cases, we explore parameter grids around this local best fit to see if there are other minima. The results show that the local best fit is the only minimum in nearby parameter space for coplanar edge-on orbits. The best fit is unstable after several hundred years and dynamical stability analysis in the grids also shows that all considered fits in coplanar edge-on  orbits are unstable after a short time. Based on the edge-on best fit, we vary the inclination to explore inclined coplanar fits. Unlike the 2:1 MMR  and 3-planet cases, the $\chi^2_{\nu}$ of  coplanar 1:1 resonance fits do not improve when the orbits are allow to be inclined, and we do not find any stable fit when all inclinations are explored. The results are similar to that of \citet{Gozdziewski2006}, who did not find any stable fit for {\it coplanar} 1:1 resonance orbits.

\subsubsection{Mutually Inclined Fits}
Next we allow the orbits to be mutually inclined and adopt the best-fit model with a coplanar edge-on 1:1 resonant configuration as an initial guess. The LM algorithm found a local minimum with $\chi^2_{\nu}$ of 1.51 and rms of $4.72\mps$. This fit is reported as fit (a), and the fitting parameters are listed in Table \ref{tab:best11}.  The mutual inclination of the orbits in fit (a) is about $78^{\circ}$, and the fit becomes unstable quickly in hundreds of years. A grid search based on fit (a) does not find any better fit in nearby parameter space and dynamical analysis  does not find any stable fit in these grids. 

\begin{figure}
\centering 
\includegraphics[width=0.95\textwidth]{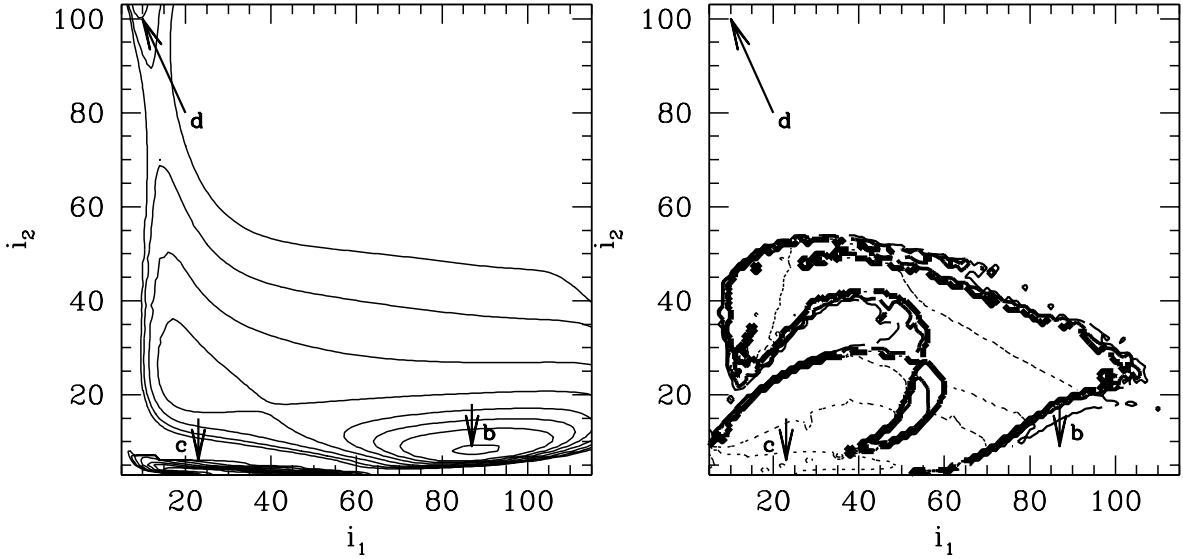}
\caption{ $\chi^2_{\nu}$ contours and dynamical properties in the $i_1-i_2$ grid for mutually inclined 1:1 resonance fits. The left panel shows $\chi^2_{\nu}$ contours (1.5, 1.55, 1.6, 1.64, 1.67, 1.7,  1.75, 1.8, 1.84)  in the $i_1-i_2$ grid, and the fits b, c, and d correspond to three $\chi^2$ minima (minimum ``d'' is not actually in the grid but is recognized from the tendency of the  $\chi^2_{\nu}$ contours). In the right panel,  the thick dashed lines are the dynamical stability boundary, and the thin dot lines  represent the contours of libration amplitudes $(25^{\circ}, 40^{\circ}, 50^{\circ}, 80^{\circ})$ of the resonance angle $\theta= \lambda_1-\lambda_2$. Solution ``c" is the only stable $\chi^2$ minimum, but this is found to have a much higher $\chi^2_\nu$ than the 2:1 MMR coplanar best fit, hence the 1:1 resonance model is not preferred.
\label{fig:i1i211}
}
\end{figure}

\begin{figure}
\centering 
\includegraphics[width=0.95\textwidth]{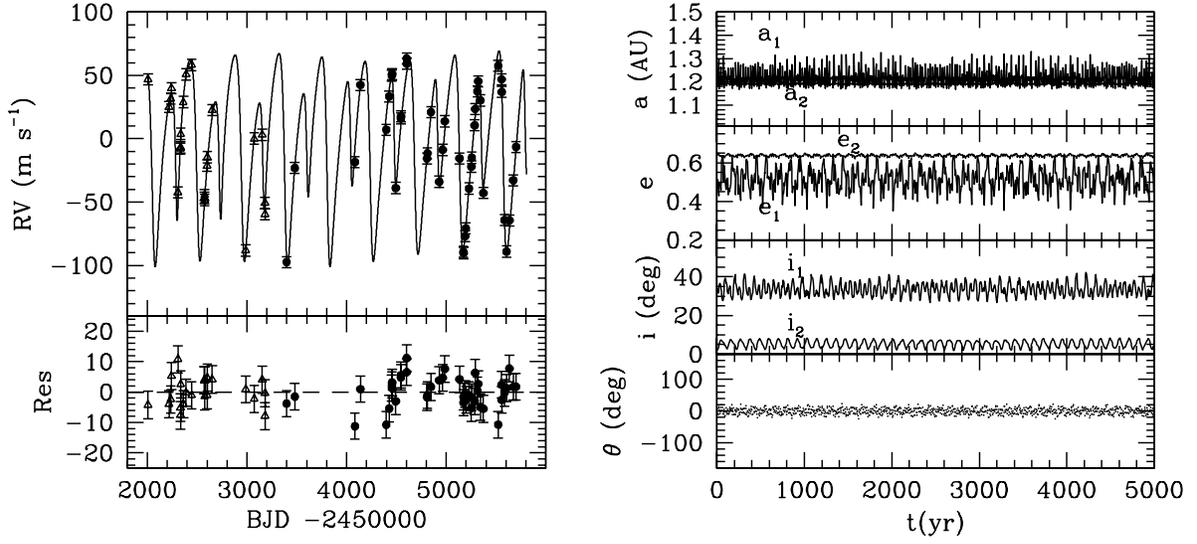}
\caption{ RV curve and residuals of the 1:1 resonance mutually inclined fit (c) in Fig. \ref{fig:i1i211} and its dynamical evolution. In the right panel, $\theta=\lambda_1-\lambda_2$. Interestingly, this 1:1 resonance model with large planetary masses ($m_1 \sim 6.1 M_{\rm{J}} ,m_2 \sim  37 M_{\rm{J}}$) remains long-term stable.
\label{fig:11}
}
\end{figure}

However, the exploration of mutually inclined fits in \citet{Gozdziewski2006} shows that for large mutual inclinations there may be multiple minima present in the parameter space, and  some of the fits with large mutual longitude of ascending node ($\Delta\Omega \sim 150^{\circ}$) are stable. We adopt the mutually inclined stable best fit of \citet{Gozdziewski2006} as the initial guess for another exploration. After we adjusted the mean anomaly $\mathcal{M}$ of their best fit to our initial epoch, we find a local minimum with $\chi^2_{\nu}$ of 1.61 and rms of $4.87\mps$ and it is reported as fit (b), with parameters  listed in Table \ref{tab:best11}. Fit (b) is highly mutually inclined with the mutual inclination of about $84^{\circ}$, and becomes unstable in a few hundred years. However, grid search based on fit (b) yields other $\chi^2$ minima. Here we show a representative $i_1-i_2$ grid which starts from the fit (b) and yield  two other minima in the grid. The left panel of Fig. \ref{fig:i1i211} shows the  $\chi^2_{\nu}$ contours $(1.50\sim1.84)$, with the arrows pointing  to the locations of three (potential) minima. The minimum in the lower left grid is labeled as fit (c) in the figure, whose parameters are listed in Table \ref{tab:best11}. The  $\chi^2_{\nu}$ of 1.59 of fit (c) is slightly less than fit (b), and Fig. \ref{fig:11} shows the RV curve and residuals of fit (c) as well as its dynamical evolution.  Interestingly, fit (c) is dynamically stable with the mass of planet 2 being about 37 times that of Jupiter,  and with  a small libration amplitude of $\theta=\lambda_1-\lambda_2$ as shown in the right panel of Fig. \ref{fig:11}.  Finally, the fit (d) which is pointed out in the left upper corner of Fig. \ref{fig:i1i211} is not actually located in the grid, but is recognized from the tendency of the contour directions. We take a fit in the region of the grid where arrow (d) is pointing, and allow all parameters to vary.  A local minimum is then found and it is reported as fit (d), with its parameters listed in Table \ref{tab:best11}. Its  $\chi^2_{\nu}$ of 1.43 is the lowest among all best fits of 1:1 resonance, however, the orbital configuration of fit (d) is retrograde with the mutual inclination of about $140^{\circ}$, and is unstable in less than a hundred years. The right panel of Fig. \ref{fig:i1i211} shows the dynamical analysis in  the 
$i_1-i_2$ grid. The thick  lines are the dynamical stability boundary, and the stable region is the region with thin dashed lines. The thin dashed lines represent the libration amplitude of $\theta=\lambda_1-\lambda_2$. All stable fits are in 1:1 resonance with $\theta$ librating around $0^{\circ}$.  Surprisingly, fits in the lower left corner of the grid, whose masses are much larger than Jupiter mass, are in a stable 1:1 resonance, whereas fits at high inclinations are unstable. 

In summary for the 1:1 resonance fits, five local best fits have been found for 1:1 resonance, and  stable 1:1 resonance fits have been found by grid search. However, the lowest $\chi^2_{\nu}$  among all 1:1 fits is 1.43, which is still much larger than 1.08 from the 2:1 MMR best fit. Therefore, there is no evidence supporting the 1:1 resonance model for the HD 82943 system.

\section{ERROR ESTIMATION}
Error estimation is important because it provides an evaluation of the uncertainties in the planetary masses and orbital parameters in the best-fit model. Based on the $\chi^2$ analysis of \S 4, the coplanar $20^{\circ}$ inclined 2:1 MMR best fit is adopted as the orbital solution for the HD 82943 planetary system. Here we analyze the errors and distribution of fitting parameters for this best fit  based on  the bootstrap  method.

 The prescription in \citet{Press1992} is adopted as our bootstrap method for fitting parameter distribution estimation\footnote{This is different from the bootstrap procedure often used in error estimation of orbital elements for exoplanets  in which the residuals from the best fit are scrambled instead of the data themselves (e.g., \citealp{Wang2012}).}. The bootstrap method uses the actual data set $\mathcal{D}$ containing $N$ data points to generate synthetic data sets $\mathcal{D}_1,\mathcal{D}_2,\ldots,$ also with $N$ data points. Here, each synthetic data set consists of 64 entries, and each data entry is chosen randomly from all 64 entries in the real data set $\mathcal{D}$ (Table \ref{tab:RV}). Each entry includes the observational time, radial velocity and instrumental uncertainty. Because of the random process, the synthetic data set almost certainly contains duplicated data points, i.e., they have the same observational time, radial velocity and uncertainty. For convenience, in the procedure of generating synthetic data set, when an entry of the real data set is chosen more than once, a random number of absolute value $\sim |0.001|$ days is added to the observational time for every  duplicated data point in the synthetic data set.  We generate and fit 5000 samples to estimate the distribution of fitting parameters and calculate the  68.3 percent confidence interval for the model parameters by:
\begin{equation}
\int_{-\infty}^{\xi_1} \! f(x) \, \mathrm{d} x = \int_{\xi_2}^{\infty} \! f(x) \, \mathrm{d} x = \frac{1-0.683}{2},
\end{equation}
where $f(x)$ is the probability density function as a function of $x$, and $\xi_1$, $\xi_2$ are the lower and upper value of 68.3 percent confidence errors, respectively. 
%As will be shown, most distributions of our best-fit parameters are nearly symmetrically distributed, so this formula is adequate for estimating the uncertainties in our case.

\begin{figure}
\centering 
\includegraphics[width=0.8\textwidth]{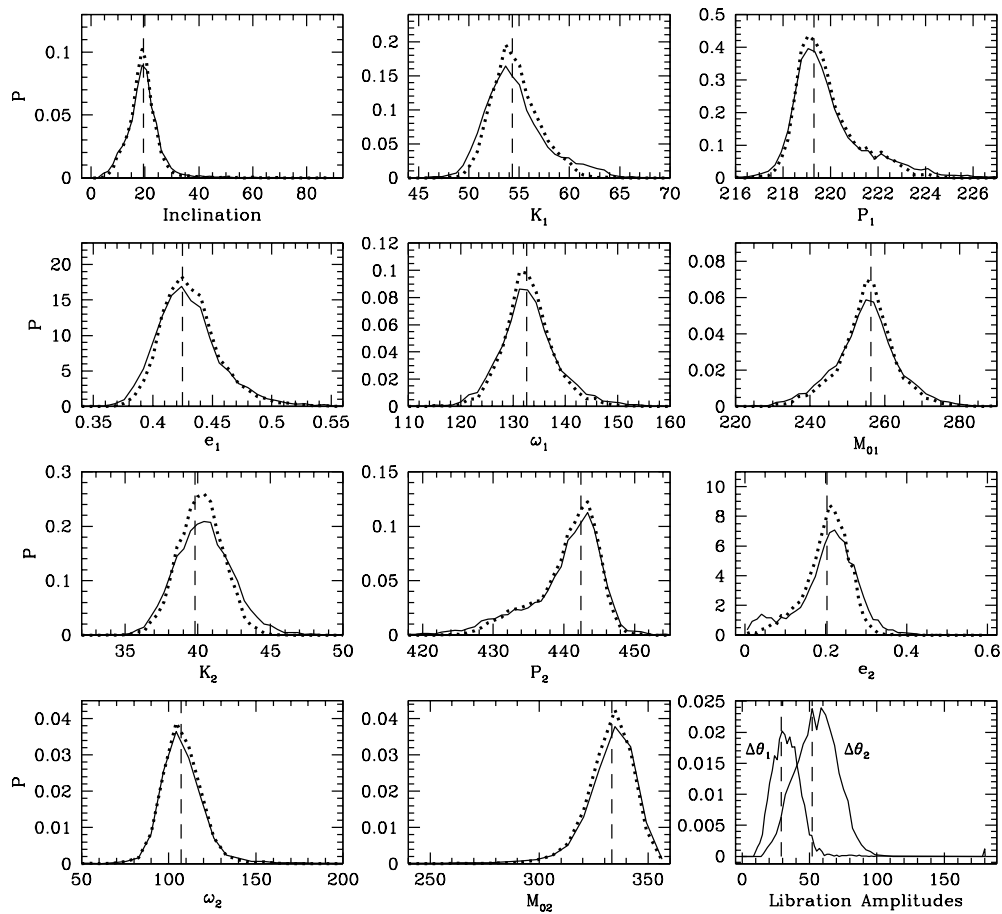}
\caption{ Probability density distribution of fitting parameters for coplanar 2:1 MMR model from bootstrap.  The solid curves are distributions of fitting parameters from all fits in bootstrap samples, the dotted curves are distributions from only  dynamically stable fits in bootstrap samples, and the dashed lines in vertical direction represent the coplanar $20^{\circ}$ inclined best-fit parameters.  The peaks of the distributions of e.g. $K_1$ and $e_2$ are slightly shifted relative to the best-fit values, suggesting that these parameters are particularly sensitive to certain data points in the original data set. Distributions of only dynamically stable fits are not significantly different from distributions from full samples, but they are more centrally peaked than those from full samples. In particular, the small peak in $e_2\approx0.05$ vanishes after stability test, suggesting that the orbital configurations with small $e_2$ are not preferred. The lower right panel is the distribution of libration amplitudes of $\theta_1$ and $\theta_2$ from dynamically stable fits in bootstrap. Dashed lines represent values from the coplanar $20^{\circ}$ inclined best fit. The distributions of both libration angles peaks near the best-fit values, favoring moderate libration amplitudes for both $\theta_1$ and $\theta_2$.
\label{fig:bootifree}
}
\end{figure}

The probability density distribution of fitting parameters determined by the bootstrap method is illustrated in Fig. \ref{fig:bootifree}. In the figure, the solid curves are the probability density distributions from all fits in bootstrap samples and the dashed lines in vertical direction represent the best-fit parameters. As shown in the figure, most  parameters are centrally peaked, and some of them are asymmetrically distributed. Some parameters' distributions do not peak at the best-fit parameters. 
For example, the peak of the distributions of $K_1$ and $e_2$ are slightly shifted from the best-fit parameters, hinting that these fitting parameters may be sensitive to some data points in the original data set. When some of the sensitive data points are absent from the synthetic data sets, the fitting results are slightly shifted from that of the original data set. Interestingly, the distribution of $e_2$ has double peaks, with a smaller one near about 0.05, hinting a small probability  for small initial $e_2$ orbital configurations.
All fits from bootstrap are integrated for a maximum time of 50,000 yr in order to examine whether dynamical stability will provide any constraints on the parameter distributions. As shown in Fig. \ref{fig:bootifree}, the dotted curves are probability density distributions from only the {\it stable} fits in bootstrap samples. The probability density distributions from the stable fits are similar to  those from all  fits, but almost all distributions from stable fits are more centrally peaked than those from all fits, meaning that although dynamical stability does not prefer a significantly different distribution, it does constrain the parameters better. In particular, the distributions of $e_2$ show a difference for small eccentricities. The small peak at $e_2\approx0.05$ vanishes after dynamical stability constraints. Thus the possibility of orbital configurations with small initial $e_2$ is ruled out by dynamical stability test. Finally, all stable fits are in 2:1 MMR with at least  $\theta_1$ librating about $0^{\circ}$ as shown in the lower right panel of Fig. \ref{fig:bootifree}, where there are a few cases with $\theta_2$ circulating.  The distributions of the libration amplitudes of both angles peak near the values of the best fit, and the overall range favor moderate libration amplitudes for both $\theta_1$ and $\theta_2$. 

Finally, the uncertainties in the orbital parameters for the 2:1 MMR coplanar best fit are listed in Table \ref{tab:error}, as determined by three methods: the covariance matrix $\sqrt{C_{ll}}$, the constant $\Delta\chi^2$ method. The inclination $i$ of the orbits is well constrained at $20^{+4.9}_{-5.5}$ degrees.   Uncertainties determined by the bootstrap method are suggested as the reported uncertainties for the best-fit parameters of this system.  The intervals of $1\sigma$ error bars determined by the constant $\Delta\chi^2$ method are comparable to those obtained from the covariance matrix. The errors from bootstrap are the largest amongst the three in every orbital fitting parameters. Both constant $\Delta\chi^2$ and bootstrap methods show asymmetric errors.

\section{COMPARISONS WITH SELF-CONSISTENT BAYESIAN ANALYSIS}

\subsection{Bayesian (DEMCMC) Approach}
\label{MCMC:METHOD}
We also analyzed the radial velocity measurements using a Bayesian framework following \citet{Ford2005} and \citet{Ford2006}. 
We assume priors that are uniform in the logarithm of orbital period, eccentricity, argument of pericenter, mean anomaly at epoch, and the velocity zero-point. 
For the velocity amplitude ($K$) and jitter ($\sigma_j$), we adopted a prior of the form
$p(x) = (x + x_o)^{-1} [\log(1 + x/x_o)]^{-1}$
with $K_o = \sigma_{j,o} = 1$ m s$^{-1}$, i.e., high values are penalized
(for a discussion of priors, see \citealt{Ford07c}). 
The likelihood for radial velocity terms assumes that each radial velocity observation is independent and normally distributed
about the true radial velocity with a variance of
$\sigma_i^2 + \sigma_j^2$, 
where 
$\sigma_i$
is the  internal measurement uncertainty, and 
$\sigma_j$
is the jitter parameter.

We used an MCMC method based upon Keplerian orbit fitting to calculate a sample from the posterior distribution \citep{Ford2006}. 
We calculated multiple Markov chains, each with 
$\sim 2 \times 10^8$
states, and discarded the first half of the chains. 
We calculated Gelman-Rubin test statistics for each model parameter and several ancillary variables and found no indications of non-convergence amongst the individual chains.
Finding no indications of non-convergence, we randomly choose a subsample ($\sim 25, 000$) from the posterior distribution for further investigation.

Following the Keplerian fitting procedure, we use the method described in \citet{Payne2011}, \citet{Johnson2011} and \citet{Wang2012}, using the subsample as the basis for a much
more computationally demanding analysis that uses fully self-consistent $N$-body integrations to account for planet-planet interactions when modeling the RV observations.
We again perform a Bayesian analysis, but replace the standard MCMC algorithm with a Differential Evolution Markov chain Monte Carlo (DEMCMC) algorithm \citep{TerBraak2006, Veras09, Veras10}. 
In the DEMCMC algorithm each state of the Markov chain is an ensemble of orbital solutions. 
The candidate transition probability function is based on the orbital parameters in
the current ensemble, allowing the DEMCMC algorithm to sample more efficiently from high-dimensional parameter spaces that have strong correlations between model
parameters.
The priors for the model parameters are the same as
those of the MCMC simulations.

For the $N$-body integrations, we use a time symmetric 4th order Hermite integrator that has been optimized for planetary systems \citep{Kokubo1998}. 
We extract the radial velocity of the star (in the barycentric frame) at each of the observation times for comparison to RV
data. 
During the DEMCMC analysis, we also impose the constraint of short-term (100 years) orbital stability.
We check whether the planetary semimajor axes remain within a factor of $50\%$ of their starting value, and that no close-approaches occur within 0.1 times the semimajor axis during the 100-year $N$-body integration. 
Any systems failing these tests are rejected as unstable (regardless of the quality of the fit to RV data). 
Thus, the DEMCMC simulations avoid orbital solutions that are violently unstable. 
In our DEMCMC simulations, this process is repeated for 10,000 generations, each of which contains $25,000$ systems, for a total of $\sim 2.5\times 10^8$ $N$-body integrations in each DEMCMC simulation.

Due to the very high computational cost of running large number of
$N$-body integrations, we confine the majority of our DEMCMC
investigations to the \emph{coplanar fixed-inclination} regime (with
$i=20^\circ$ or $90^\circ$).  We leave DEMCMC analysis of fits that
allow the inclination to vary for future work.

\subsection{Results of MCMC RV Analysis}

We now present the results of our application of the MCMC methodology described in \S  \ref{MCMC:METHOD} to the RV data sets listed in Table \ref{tab:RV}, and compare the results to those from the $\chi^2$ minimization with grid search and the bootstrap method. 

\subsubsection{2-Planet Keplerian Fits }\label{RES:2:KEP}

We illustrate a sample of the Keplerian MCMC analysis by plotting in Fig.  \ref{FIG:RES:NB:1}  the periods of the two planets, $(P_1,P_2)$ which result from analyzing the Keck  data sets. 
We show contour plots for the planetary periods ($P_1,P_2$) for the frequentist Levenberg-Marquardt (thin, black, solid) and Bayesian MCMC (thick, black, dotted) approaches. 
The thin solid black contours display the different $\chi^2$ levels (the minimum $\chi^2$ and $1,2 \sigma$ according to $\Delta \chi^2$ from inside out). 
The thick dotted black contours display \emph{different} information, as the MCMC algorithm provides us with a final density of solutions, so we plot iso-density contours  containing $25\%$, $68.2\%$ and $95.4\%$ of solutions. 
The 2:1 period ratio is plotted as a gray dotted line. 
We find that the two approaches agree very well,  in the sense  that the size and shape of contours are very similar and they both find \emph{consistent} best-fit solutions very close to the 2:1 period ratio. 

The probability density distribution of fitting parameters  resulting from the Bayesian MCMC approach is plotted in Fig. \ref{fig:bootmcmc} as the red curves, and the vertical lines represent the best-fit values from $\chi^2$ minimization. The majority of the parameters  display smooth Gaussian profiles, and most of the peaks nearly coincide with the best-fit values. It should be noted that the eccentricity of the outer planet is rather poorly constrained in this Keplerian analysis. 
Using the bootstrap method described in \S 5, we generate 5,000 synthetic data sets and fit them with a Keplerian model. The distribution from bootstrap is shown as the black curves in Fig. \ref{fig:bootmcmc}, and many of them show less constraints (e.g., $K_1, e_1, m_{\rm{min1}}, P_2, \omega_2$) than the distribution from the MCMC approach. 
A full table of the mean values and their $1\sigma$ errors from both MCMC and bootstrap is given in Table  \ref{tab:error4}. All the uncertainties from bootstrap are larger than those from the MCMC approach.

\subsubsection{2-Planet Coplanar $20^{\circ}$ Inclined Fits: $N$-Body}\label{RES:2:NBODY}

\begin{figure}
%\begin{minipage}[b]{\textwidth}
\centering
\includegraphics[angle=-90,width=0.8\columnwidth]{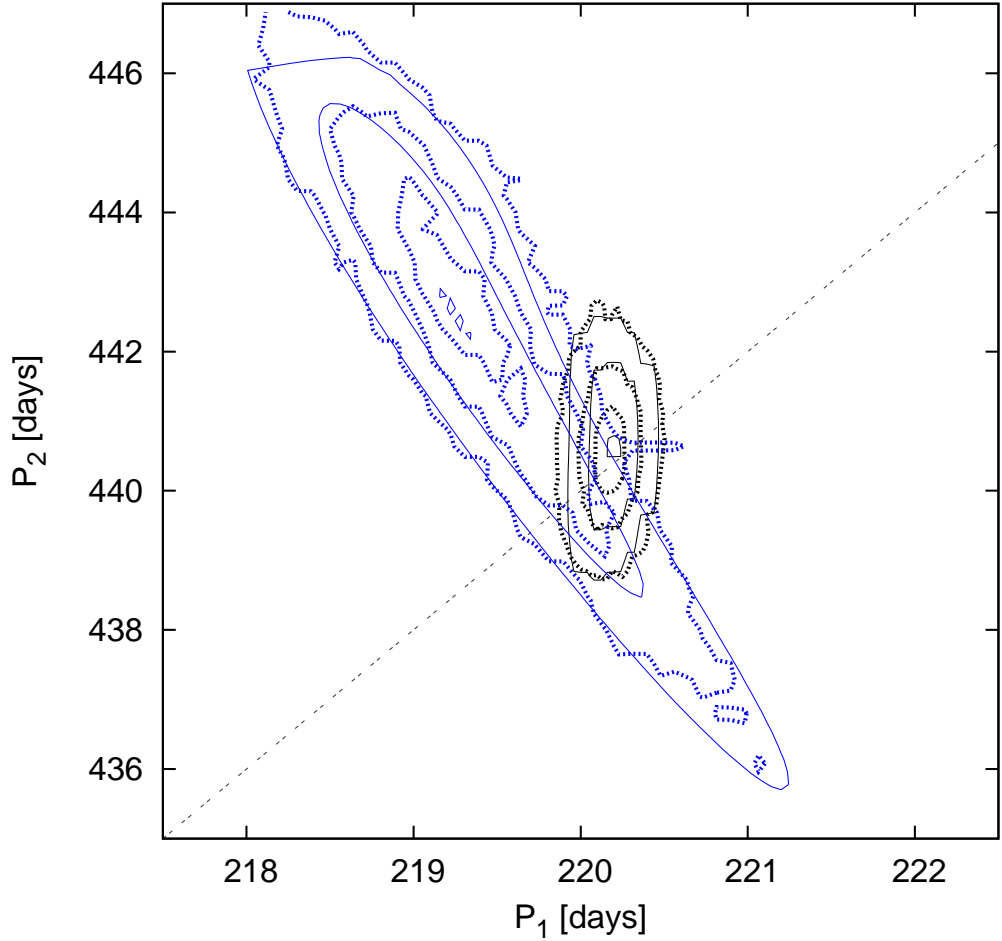}\\
\caption{
Contour plots for the planetary periods ($P_1$, $P_2$) for the
frequentist Levenberg-Marquardt (thin, solid lines) and Bayesian
(thick, dotted lines) approaches.
The black lines represent the Keplerian fits, and
%the red lines at the lower right are the results of the N-body Þtting procedures for $i=90^{\circ}$, 
the blue lines are the results of the $N$-body fitting procedures for $i=20^{\circ}$. 
For the Levenberg-Marquardt fits, the three contours display different
$\chi^2_{\nu}$ levels (the minimum $\chi^2_{\nu}$, $1\sigma$,
$2\sigma$ from inside out), while for the Bayesian fits, the
iso-density contours contain 25\%, 68.2\% and 95.4\% of the (DE)MCMC
solutions.
The 2:1 period ratio is plotted as a gray dotted line. 
Results from both methods show excellent agreement with each other.
%Contour plots for the planetary periods ($P_1,P_2$) for the frequentist Levenberg-Marquardt (black) and Bayesian (red) approaches. The thick lines in upper left of the plot represent the Keplerian fits, and the thin lines in lower right of the plot are the results of the $N$-body fitting procedures. 
%
%For Keplerian fits (upper left), the thick dotted black contours display the different $\chi^2$ levels (the minimum $\chi^2$, $1\sigma, 2\sigma$ from inside  out), and the thick dashed red contours display the iso-density contours containing $25\%$, $68.2\%$ and $95.4\%$ of MCMC solutions. The 2:1 period ratio is plotted as a gray dotted line.
%For $N$-body fits (lower right), the thin solid black contours display the different $\chi^2$ levels (the minimum $\chi^2$, $1\sigma, 2\sigma$ from inside out) for the Levenberg-Marquardt fits, and the thin dashed red lines are the iso-density contours of $25\%$, $68.2\%$, and $95.4\%$ of DEMCMC solutions.
}
%\end{minipage}`
\label{FIG:RES:NB:1}
\end{figure}

\begin{figure}
\centering
\includegraphics[width=0.8\columnwidth]{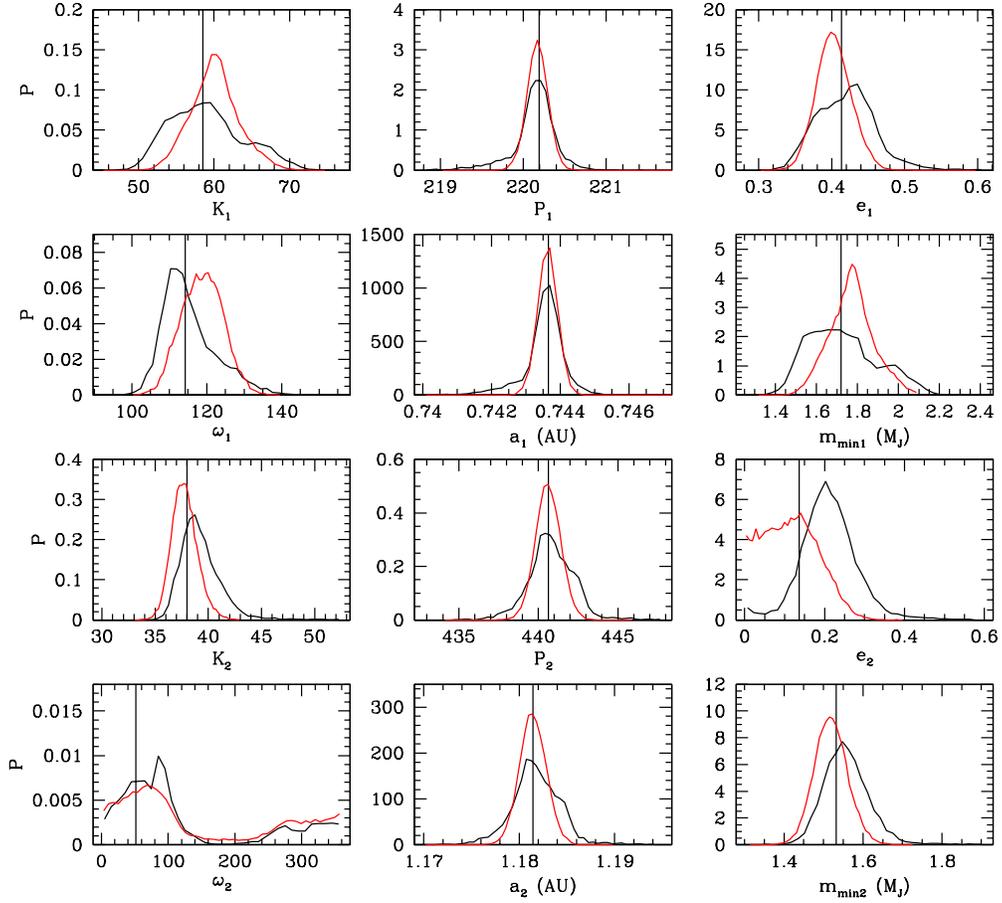}
\caption{Probability density distributions from bootstrap (black curves) and the MCMC approach (red curves) for Keplerian fits.  The vertical lines represent the best-fit values from $\chi^2$ minimization. For the MCMC results, the majority of the parameters display smooth gaussian profiles, and most of the peaks coincide with the best-fit values. The distribution from bootstrap show less constraints on parameter than the MCMC approach.
}
\label{fig:bootmcmc}
\end{figure}

Implementing the $N$-body fitting procedures detailed in \S
\ref{MCMC:METHOD}, we arrive at the parameter fits illustrated by the
blue contours in Fig. \ref{FIG:RES:NB:1} for $i = 20^\circ$, with
similar contents as the Keplerian fits.
We find in Fig. \ref{FIG:RES:NB:1} that the inclusion of mutual
interactions leads to the period of the inner planet (at the first
observing epoch) shifting to a slightly shorter period (from $P_1\sim
220$ days, to $P_1\sim 219.2$ days), while the outer period shifts to
slightly larger values (from $P_2 \sim 440.5$ days, to $P_2 \sim
442.5$ days).
Similar to Keplerian fits, the LM and DEMCMC approaches agree very
well for $20^{\circ}$ inclined $N$-body fitting results.
The contours from $N$-body fits are larger than those from Keplerian
fits, suggesting that the Keplerian fits constrain the parameters
better than the $20^{\circ}$ inclined $N$-body fits.

\begin{figure}
\centering
\includegraphics[width=0.8\columnwidth]{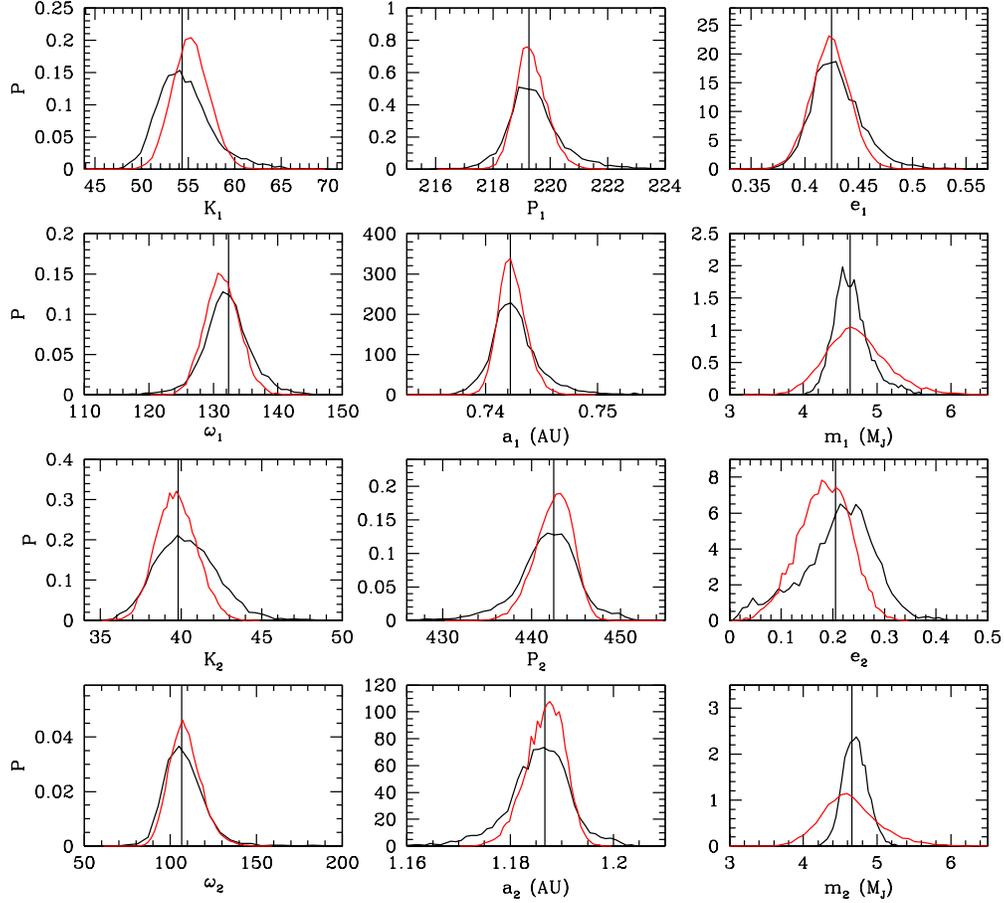}
\caption{Probability density distributions from bootstrap (black
curves) and DEMCMC (red curves) for coplanar $i = 20^{\circ}$
dynamical fits. Almost all distributions from both methods show smooth
gaussian profiles and all of them peak around the best-fit values. The
LM and DEMCMC approaches agree well for dynamical fitting results.
}
\label{fig:bootdemcmc}
\end{figure}

We also compare the results from DEMCMC to those from bootstrap, as
illustrated in Fig. \ref{fig:bootdemcmc}, where the vertical lines
represent the best-fit values from $\chi^2$ minimization from a
coplanar $20^{\circ}$ inclined dynamical model, the red solid curves
are from DEMCMC, and the black solid curves are from bootstrap. Almost
all distributions from both methods show smooth gaussian profiles, and
all of them peak around the best-fit values. Distributions from
DEMCMC and bootstrap are similar in terms of both their shapes and
sizes. The mean values and their errors are listed in Table
\ref{tab:error4}. The stellar jitter is treated as an unknown
parameter in Bayesian analysis. The distribution of jitter (not
plotted) peaks at about $4.50 \mps$ with uncertainty of $(+0.51,
-0.47)\mps$, which is consistent with the estimated value $4.2\mps$
used in \S4 and \S5.

We also provide in Table \ref{tab:error4} the mean values and errors
of fitting parameters from both DEMCMC and bootstrap for coplanar
edge-on ($i = 90^\circ$) models, which are not discussed above, as a
reference for future study. Our DEMCMC Bayesian algorithms have been
tested in ever greater detail and found to perform successfully, with
the strong overlap in the result from the different methods giving
increased confidence in the robustness of our conclusions.

\subsection{3-Planet Coplanar Edge-on Fits}\label{RES:3}
In \S\ref{3planet} we found using the periodograms and LM analysis that there was little evidence for a third planet. 
However, for the sake of completeness, we now model the system with three planets using the same MCMC approach (Keplerian and $N$-body) outlined in \S \ref{MCMC:METHOD}.

We find that the best fit solution (not shown) has the inner two planets essentially unaltered at $P_1 \sim 220$ and $P_2\sim 440$ days, with the third planet at $P_3\sim1,100$ days (although this is extremely poorly constrained, with significant uncertainties $>1,000$ days), i.e., results that are very similar to those found using the LM approach. 
Hence we are confident that there is no evidence of the 1:2:4 resonance (requiring $P_3\sim880$ days) suggested in \citet{Beauge2008}.

% ----------------------------------------------------------------------------------------------------------------

\section{SUMMARY AND DISCUSSION}

We have analyzed the orbital and dynamical state of the HD 82943 planetary system by dynamically fitting 10 years of Keck RV measurements. Based on parameter grid search, fits around the best fits as a function of various pairs of parameters have been systematically explored.  Three type of fits associated with qualitatively different orbital configurations, the 2:1 MMR, 3-planet and 1:1 resonance configurations, have been examined. 

In terms of the 2:1 MMR fits, our Keplerian best fit has $\chi^2_{\nu}$ = 1.38, significantly better than previous results ($\chi^2_{\nu} \gtrsim 1.87$) which were based on only 3.8 years of Keck RV data combined with the lower quality CORALIE data. The dynamical best fit of the coplanar edge-on orbits has $\chi^2_{\nu}$ = 1.21 and rms of  $4.37 \mps$, and it is in a 2:1 MMR with both resonance angles $\theta_1 = \lambda_1 - 2\lambda_2 + \varpi_1$ and $\theta_2 = \lambda_1 - 2\lambda_2 + \varpi_2  $ librating around $0^{\circ}$. Grid search coupled with dynamical stability test in the $h_2-k_2$ and $P_1-P_2$ grids shows that the best fit is the only $\chi^2$ minimum. The best fit is deep in the stable region and all fits in the stable region are in 2:1 MMR. When the inclination of coplanar orbits is varied,  the $\chi^2$ as a function of inclination clearly shows a deep minimum at about $20^{\circ}$, with $\chi^2_{\nu}$ of 1.08 and rms of $4.09\mps$, which is close to $3\sigma$ confidently better than the edge-on best fit.  The $20^{\circ}$ inclined best fit contains two planets of masses 4.78 and 4.80 $M_{\rm{J}}$, and it is in 2:1 MMR with both $\theta_1$ and $\theta_2$ librating around $0^{\circ}$. 
Systematic search for fits allowing the inclination to vary  in $h_2-k_2$ and $P_1-P_2$ grids shows that the best fit is also the only $\chi^2$ minimum.  All good fits are in the stable region and all stable fits are in 2:1 MMR.   The $\chi^2$ contours and dynamical properties of fits in the grids are similar to that of coplanar edge-on fits, except that  the $\chi^2$ contours in the grids show discontinuities. Finally, the mutual inclination of 2:1 MMR fits cannot be constrained by either $\chi^2$ statistics or dynamical stability test.
Compared to previous fitting results of the 2:1 MMR configuration based on the lower-quality CORALIE and shorter Keck  RV data \citep{Lee2006,Gozdziewski2006,Beauge2008}, our 2:1 MMR best-fit model  improves significantly in both $\chi^2$ and the rms. More importantly, assuming coplanar configuration, the inclination relative to the sky plane is well contained at about $20^{\circ}$, and the system is stable in a  2:1 MMR configuration.

The periodograms  of the residuals from both coplanar edge-on and $20^{\circ}$ inclined 2:1 MMR best fits do not show significant evidence for the existence of a third planet,  contrary to previous results in \citet{Beauge2008} where the periodogram of the residuals of their 2-planet best fit showed significant signal for the existence of a third planet.   When we fit for a third planet, the best fit has  $\chi^2_{\nu}$ of 0.59 and rms of $2.91\mps$. The fact that the $\chi^2_{\nu}$ is significantly lower than 1.0 and the rms is significantly lower than the estimated stellar jitter of $4.2\mps$ hints that the 3-planet model  over fits  the current  RV data. The best-fit model becomes unstable within hundreds of years due to the large $e_3$. In the $h_3-k_3$ grid, two other local $\chi^2$ minima have been found, but they are unstable. Dynamical stability test in the grid shows that only fits with $e_3 \lesssim 0.06$ remain stable for 50,000 yr. Fits in $P_2-P_3/P_2$ grid have been explored, and we did not find any good fits associated with the  Laplace resonance configuration.  
 
For the 1:1 resonance configuration, only one $\chi^2$ minimum has been found in coplanar edge-on orbits, and  several minima have been found in the mutually inclined fits. All coplanar fits and most of the mutually inclined fits are unstable, but we have found some stable fits with high mutual inclination in the $i_1-i_2$ grid. Only one minimum is stable with a small libration amplitude of $\theta= \lambda_1-\lambda_2$, and dynamical behaviors of the stable fits are similar to the stable fits found by \citet{Gozdziewski2006}. However, all fits we found have significantly higher $\chi^2_{\nu}$ ($\gtrsim 1.43$) than the 2:1 MMR best fit, so they are ruled out.

In summary, based on the $\chi^2$ statistics and dynamical stability
constraints, the 2:1 MMR coplanar $20^{\circ}$ inclined best fit is
reported as the best fit for the HD 82943 planetary system. The
best-fit parameters are listed in Table \ref{tab:error}, and their
uncertainties determined by bootstrap in Table \ref{tab:error} are
suggested as the reported uncertainties. There is no evidence for
either the 3-planet Laplace resonance fits or the 1:1 resonance
fits. The HD 82943 planetary system contains two planets in 2:1
mean-motion resonance,  and its dynamical state  is well
established. The resonant angles $\theta_1$ and $\theta_2$  of two
nearly equal-mass planets are librating around $0^{\circ}$ with
moderate libration amplitudes of about $30^{\circ}$ and $52^{\circ}$,
respectively.

 It is interesting to show the differences between the Keplerian best fit, dynamical coplanar edge-on best fit and coplanar $20^{\circ}$ inclined best fit of 2:1 MMR in graphical form using radial velocity plots, rather than $\chi^2_{\nu}$, so that one can have an  intuitive evaluation of the improvements of the fitting. More importantly, if there are significant variations  of the RV values from different fits after the last observed epoch of the Keck data sets,  RV observations in the near future may provide more constraints on our best fit.  A convenient method is to compare the residuals of two fits. For example,  we plot the residuals of  fit(a): ${\rm RV} - {\rm fit(a)}$, and then we plot a curve of the RV values of another fit (b) which is subtracted by the fit (a): ${\rm fit(b)} - {\rm fit(a)}$. By evaluating how the curve fits the residuals compared to the zero line, we can know how fit(b) improves the fitting compared to fit(a).

\begin{figure}
\centering 
\includegraphics[width=0.9\textwidth]{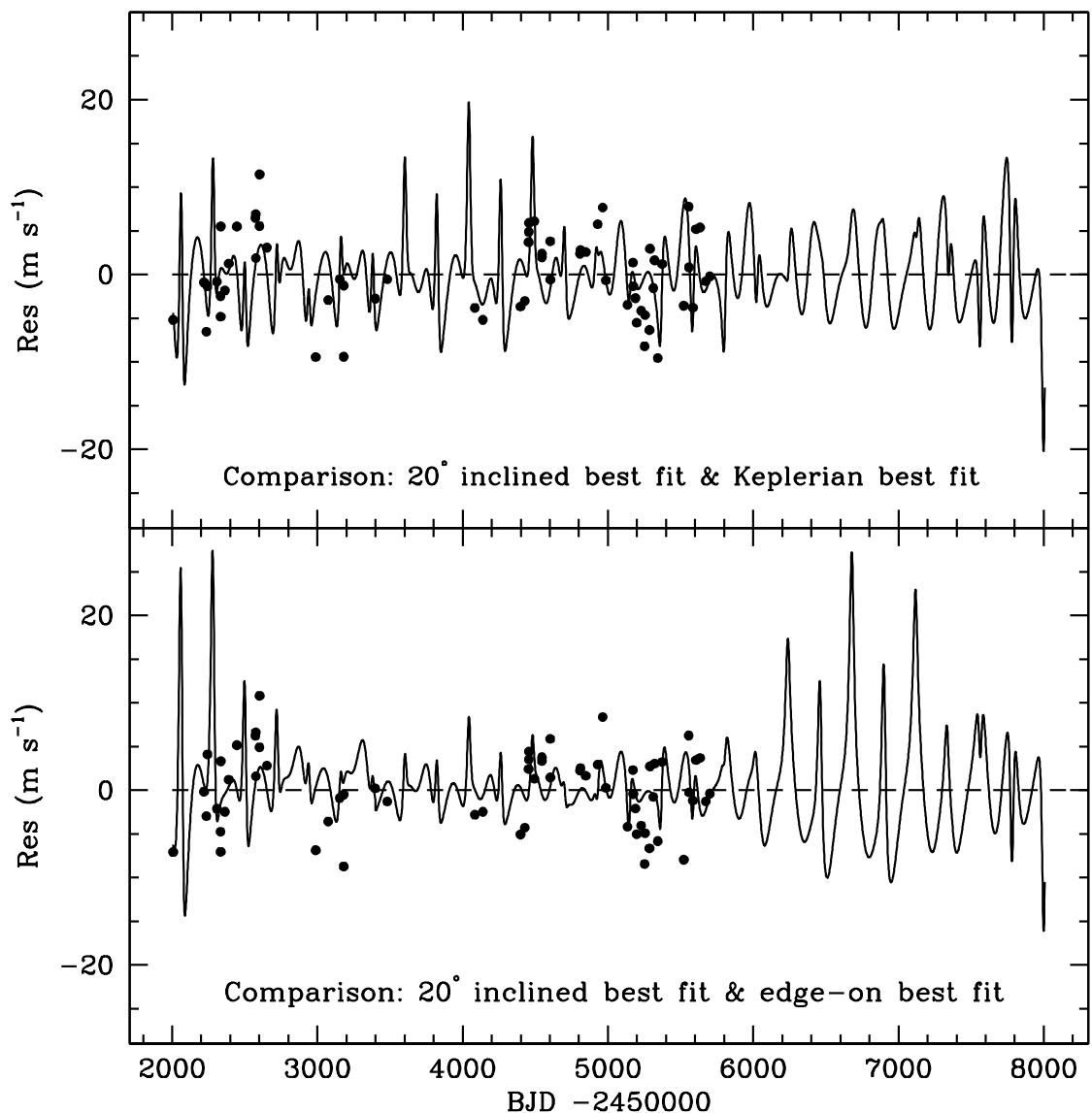}
\caption{ Upper panel: dots are the residuals of the Keplerian best fit and the curve represents the RV values of [fit($20^{\circ}$)$-$fit(Kep)]. The fluctuations show NO systematic trend in the observational timescale, suggesting the improvements to the fit primarily stem from the short-term mutual interactions between the planets, instead of precession of orbital periapses.
Lower panel: dots are the residuals of the coplanar edge-on best fit, and the curve represents the RV values of [fit($20^{\circ}$)$-$fit(edge-on)]. The large peaks around BJD 2456700 may allow future observations to better constrain the true inclinations and masses of the planets in the system.
\label{fig:pre}
}
\end{figure}

First, we compare the Keplerian best fit to the coplanar $20^{\circ}$ inclined best fit. The upper panel of Fig. \ref{fig:pre} shows the differences of the residuals from the Keplerian and the $20^{\circ}$ inclined best fit, in which the dots are the residuals of the Keplerian best fit and the curve represents the RV values of [fit($20^{\circ}$)$-$fit(Kep)]. The fluctuations of the curve oscillate around the zero line and  do not show an obvious systematic trend in the observation time span (10 yr). This situation suggest that the improvement from the Keplerian best fit to the coplanar $20^{\circ}$ inclined best fit is primarily contributed by the short term mutual interactions of the planets. A large fraction of dots are apparently fitted better by the curve than the zero line. For example, the first dot and dots around BJD  2,454,500  and BJD 2,455,500 significantly deviate from zero but they are much closer to  the curve. The orbital precession rate $\dot{\omega}$ is on the order of only about $1.5^{\circ}~ \rm{yr}^{-1}$ (or $15^{\circ}$ in 10 yr) for the coplanar $20^{\circ}$ inclined best fit, so it is reasonable that we primarily see improvement from short term interactions. The fluctuations in the near future do not show significant peaks until BJD 2,458,000.
Next we compare the coplanar edge-on best fit to the coplanar $20^{\circ}$ inclined best fit in the lower panel of Fig. \ref{fig:pre}, in the same format as the upper panel. In this comparison, the fluctuations are smaller than in the comparison associated with the Keplerian best fit in the observed time span,  except for the large peaks at the beginning.  Unlike the comparison with the Keplerian best fit, the improvements are not so obvious as a large fraction of the dots are not obviously fitted well by the curve. Interestingly, the curve shows large peaks at around BJD 2,456,800, which is about 3 yr after the last observed epoch in the Keck data. Thus future RV observations at around that time  could provide more constraints on the inclinations and the true planetary masses of the HD 82943 planetary system.

During the course of the submission and review of this article, we learned of a complimentary investigation by Kennedy et al (in press) in which Herschel observations of HD 82943 detect a debris-disk with an inner edge $>100\,AU$ (far beyond the planets studied in our analysis). The debris-disk appears to have a best-fit inclination of approximately $27\pm4$ degrees (to the plane of the sky), strongly supporting the inclination we deduced from our purely dynamical studies.

\acknowledgments 
We thank NASA and NExScI for providing Keck time in the 2011A semester
for the study of multiplanet systems (NExScI ID40/Keck ID\# N141Hr).
Data presented herein were obtained at the W. M. Keck Observatory from
telescope time allocated to the National Aeronautics and Space
Administration through the agency's scientific partnership with the
California Institute of Technology and the University of California.
The Observatory was made possible by the generous financial support of
the W. M. Keck Foundation.
We thank Howard Isaacson, R.P. Butler, and S. Vogt for help with observing at the telescope, and the referee for helpful comments on the manuscript.
X.T.\ and M.H.L.\ were supported in part by the Hong Kong RGC grant
HKU 7034/09P.
Contributions by M.J.P.\ and E.B.F.\ were supported by NASA Origins of Solar Systems grant NNX09AB35G only prior to August 8, 2011.
J.A.J.\ was supported by generous grants from the David and Lucile
Packard Foundation and the Alfred P. Sloan Foundation.
J.T.W.\ was supported by NSF Astronomy and Astrophysics Grant AST-1211441.
The Center for Exoplanets and Habitable Worlds is supported by the
Pennsylvania State University, the Eberly College of Science, and the
Pennsylvania Space Grant Consortium.
M.H.L.\ and E.B.F.\ also acknowledge the hospitality of the Isaac
Newton Institute for Mathematical Sciences, where part of this work
was completed.

%\bibliography{apj-jour,myrefs}
\bibliographystyle{apj}   
\bibliography{hd82943v1}

%%%%%%%%%% table of Keck data sets %%%%%%%%%%%%

\begin{deluxetable}{rrrrr}
\tablecolumns{5}
\tablewidth{0pt}
\tablecaption{Radial Velocities of HD 82943 from Keck 
\label{tab:RV}}
\tablehead{
\colhead{JD} & \colhead{} & \colhead{Radial Velocity} & \colhead{} &
\colhead{Uncertainty} \\
\colhead{($-$2450000)} & \colhead{} &
\colhead{(${\rm m}\,{\rm s}^{-1}$)} & \colhead{} &
\colhead{(${\rm m}\,{\rm s}^{-1}$)}
}
\startdata
\multicolumn{5}{c}{Data Set 1}\\
\noalign{\vskip .50ex}
\cline{1-5}
\noalign{\vskip .50ex}     
      2006.9130 & &      43.90  & &    1.56 \\  
      2219.1210 & &     22.08   & &   1.36 \\  
      2236.1262 & &    28.57    & &  1.34 \\  
      2243.1295 & &      36.80  & &    1.35 \\  
      2307.8391 & &     $-$45.39  & &    1.51 \\  
      2332.9834 & &     $-$9.84  & &    1.72 \\  
      2333.9558 & &     $-$10.83  & &    1.63 \\  
      2334.8726 & &     0.77 & &     1.49 \\  
      2362.9717 & &      25.84  & &    1.66 \\  
      2389.9438 & &      47.74  & &    1.57 \\  
      2445.7387 & &      55.28  & &    1.55 \\  
      2573.1473 & &     $-$49.82  & &    1.42 \\  
      2575.1400 & &     $-$47.46  & &    1.36 \\  
      2576.1437 & &     $-$51.51  & &    1.55 \\  
      2601.0664 & &     $-$24.35  & &    1.57 \\  
      2602.0728 & &    $-$17.51   & &   1.47 \\  
      2652.0009 & &      19.62  & &    1.55 \\  
      2988.1092 & &     $-$91.26  & &    1.40 \\  
      3073.9287 & &     $-$2.98  & &    1.56 \\  
      3153.7544 & &     0.00   & &   1.41 \\  
      3180.7448 & &     $-$62.77  & &    1.36 \\  
      3181.7416 & &     $-$53.85  & &    1.51 \\  
\noalign{\vskip .50ex}
\cline{1-5}
\noalign{\vskip .50ex}
\multicolumn{5}{c}{Data Set 2}\\
\noalign{\vskip .50ex}
\cline{1-5}
\noalign{\vskip .50ex}     
      3397.9083 & &       $-$107.66  & &    1.13 \\  
      3480.7572 & &       $-$33.53  & &   0.97 \\  
      4084.0897 & &     $-$28.82    & & 0.88 \\  
      4139.0165 & &      32.21    & &  1.06 \\  
      4398.1361 & &     $-$3.17    & &  1.05 \\  
      4428.1286 & &      23.07    & &  1.15 \\  
      4454.1151 & &      39.44    & & 0.96 \\  
      4455.0271  & &     40.27    & &  1.05 \\  
      4456.0837  & &     40.81    & & 0.98 \\  
      4492.9843  & &    $-$49.24    & &  1.31 \\  
      4544.9394   & &    5.61    & & 0.67 \\  
      4545.9278  & &     7.24    & & 0.67 \\  
      4602.8125  & &     52.94    & &  1.27 \\  
      4603.7880  & &     48.85    & &  1.25 \\  
      4807.1705  & &    $-$25.89    & &  1.37 \\  
      4811.1235  & &    $-$21.83    & &  1.18 \\  
      4847.0406   & &    10.63    & &  1.41 \\  
      4929.8077  & &    $-$44.36    & &  1.41 \\  
      4963.8378  & &    $-$19.10    & &  1.33 \\  
      4985.7698  & &     3.45    & &  1.21 \\  
      5134.0952  & &    $-$25.88    & &  1.20 \\  
      5172.1518  & &    $-$99.47    & &  1.19 \\  
      5174.0713  & &    $-$100.61    & &  1.13 \\  
      5189.1241  & &    $-$87.31    & &  1.10 \\  
      5197.9924   & &   $-$81.09    & &  1.26 \\  
      5229.0603   & &   $-$49.61    & &  1.31 \\  
      5252.0320   & &   $-$32.41    & &  1.41 \\  
      5255.8658  & &    $-$25.32    & &  1.23 \\  
      5285.8518  & &    0.23   & &   1.25 \\  
      5289.8818  & &     13.11    & &  1.25 \\  
      5312.7844  & &     27.11    & &  1.18 \\  
      5320.7774  & &     35.04    & &  1.13 \\  
      5342.7496  & &     19.90    & &  1.18 \\  
      5372.7395  & &    $-$53.19    & &  1.26 \\  
      5522.0604  & &     47.18    & &  1.27 \\  
      5556.0871  & &     36.55    & &  1.29 \\  
      5558.0063  & &     26.56    & &  1.13 \\  
      5585.0690  & &    $-$74.56    & &  1.32 \\  
      5605.9988  & &    $-$99.45    & &  1.39 \\  
      5633.7992  & &    $-$74.82    & &  1.26 \\  
      5672.8239  & &    $-$43.12    & &  1.20 \\  
      5700.7478  & &    $-$16.85    & &  1.27 \\  
      \enddata

\end{deluxetable}

%%%%%%%% table of  parameters of coplanar 2:1 MMR best-fits  %%%%%%%%%%%%

\begin{deluxetable}{lcccc}
 \tablecolumns{5}
 \tablewidth{0pt}
 \tablecaption{Best fits of 2:1 MMR configuration
 \label{tab:best21}}
 \tablehead{
 \colhead{Parameter}&\colhead{}&\colhead{Inner Planet}&\colhead{}&\colhead{Outer Planet}\\
}
\startdata
\multicolumn{5}{c}{ Keplerian fit: $\chi^2_{\nu} = 1.38$, $rms=4.66\mps$}\\
\noalign{\vskip .20ex}
\cline{1-5}
\noalign{\vskip .20ex}
$K~ (\mps)$ && 58.5 && 38.0 \\
$P$ (days) && 220.2 && 440.6 \\
$e$ && 0.413 && 0.136 \\
$\omega$ (deg) && 114 && 52\\
$\mathcal{M}$ (deg) && 273 && 22 \\
$m_{\rm{min}}$ ($M_{\rm{J}}$)  && 1.72 && 1.53 \\
$V_1 ~(\mps)$ && \multicolumn{3}{c}{-6.4} \\
$V_2 ~(\mps)$ && \multicolumn{3}{c}{-2.0} \\
\noalign{\vskip .20ex}
\cline{1-5}
\noalign{\vskip .20ex}
\multicolumn{5}{c}{Coplanar edge-on dynamical fit: $\chi^2_{\nu} = 1.21$, $rms=4.37\mps$}\\
\noalign{\vskip .20ex}
\cline{1-5}
\noalign{\vskip .20ex}
$K~ (\mps)$ && 57.8 && 37.8 \\
$P$ (days) && 221.0 && 438.5 \\
$e$ && 0.407 && 0.096 \\
$\omega$ (deg) && 123 && 90\\
$\mathcal{M}$ (deg) && 276 && 342 \\
$m$ ($M_{\rm{J}}$)  && 1.71 && 1.53 \\
$V_1 ~(\mps)$ && \multicolumn{3}{c}{-8.0} \\
$V_2 ~(\mps)$ && \multicolumn{3}{c}{-0.3}\\
\noalign{\vskip .20ex}
\cline{1-5}
\noalign{\vskip .20ex}
\multicolumn{5}{c}{Coplanar inclined dynamical fit: $\chi^2_{\nu} = 1.08$, $rms=4.09\mps$}\\
\noalign{\vskip .20ex}
\cline{1-5}
\noalign{\vskip .20ex}
$K~ (\mps)$ && 54.3 && 39.8 \\
$P$ (days) && 219.3 && 442.4 \\
$e$ && 0.425 && 0.203 \\
$\omega$ (deg) && 133 && 107\\
$\mathcal{M}$ (deg) && 256 && 333 \\
$m$ ($M_{\rm{J}}$)  && 4.78 && 4.80 \\
$i$ (deg) && \multicolumn{3}{c}{19.4}\\
$V_1 ~(\mps)$ && \multicolumn{3}{c}{-6.6} \\
$V_2 ~(\mps)$ && \multicolumn{3}{c}{-1.5}\\
\enddata
\end{deluxetable}

%%%%%%%%%%%%%%%%%%%%%%%%%%%%%%%%

 %%%%%%%%%%%%%%% table of parameters of 3-planet fit %%%%%%%%%%
 
 \begin{deluxetable}{lcccccc}
 \tablecolumns{7}
 \tablewidth{0pt}
 \tablecaption{Best fit of edge-on 3-planet configuration
 \label{tab:3p}}
 \tablehead{
 \colhead{Parameter}&\colhead{}&\colhead{Planet 1}&\colhead{}&\colhead{Planet 2}&\colhead{}&\colhead{Planet 3}\\
}
\startdata
\multicolumn{7}{c}{ $\chi^2_{\nu} = 0.59$, $rms=2.91\mps$}\\
\noalign{\vskip .50ex}
\cline{1-7}
\noalign{\vskip .50ex}
$K~ (\mps)$ && 58.8 && 38.1 && 6.3\\
$P$ (days) && 221.2 && 438.7 && 1077.4 \\
$e$ && 0.412 && 0.068 && 0.402 \\
$\omega$ (deg) && 116 && 89 && 27\\
$\mathcal{M}$ (deg) && 279 && 343 && 148 \\
$m$ ($M_{\rm{J}}$)  && 1.73 && 1.55 && 0.314\\
$V_1 ~(\mps)$ && \multicolumn{5}{c}{-8.4} \\
$V_2 ~(\mps)$ && \multicolumn{5}{c}{1.6}\\
\enddata
\end{deluxetable}
%%%%%%%%%%%%%%%%%%%%%%%%%%

%%%%%%%% table of parameters of coplanar edge-on 1:1 best fit %%%%%%%%%%

\begin{deluxetable}{lcccc}
 \tablecolumns{5}
 \tablewidth{0pt}
 \tablecaption{Best fit of coplanar edge-on 1:1 resonance configuration
 \label{tab:11}}
 \tablehead{
 \colhead{Parameter}&\colhead{}&\colhead{Planet 1}&\colhead{}&\colhead{Planet 2}\\
}
\startdata
\multicolumn{5}{c}{ $\chi^2_{\nu} = 1.69$, $rms=5.14\mps$}\\
\noalign{\vskip .50ex}
\cline{1-5}
\noalign{\vskip .50ex}
$K~ (\mps)$ && 92.3 && 59.7 \\
$P$ (days) && 442.0 && 439.5 \\
$e$ && 0.466 && 0.654 \\
$\omega$ (deg) && 117 && 125\\
$\mathcal{M}$ (deg) && 325 && 136 \\
$m$ ($M_{\rm{J}}$)  && 3.33 && 1.83 \\
$V_1 ~(\mps)$ && \multicolumn{3}{c}{-10.1} \\
$V_2 ~(\mps)$ && \multicolumn{3}{c}{2.2}\\
\enddata
\end{deluxetable}

%%%%%%%%   table of parameters of mutually inclined 1:1 resonance best-fits  %%%%%%%%%%%%

\begin{deluxetable}{lcccc}
 \tablecolumns{5}
 \tablewidth{0pt}
 \tablecaption{Best fits of mutually inclined 1:1 resonance configuration
 \label{tab:best11}}
 \tablehead{
 \colhead{Parameter}&\colhead{}&\colhead{Planet 1}&\colhead{}&\colhead{Planet 2}\\
}
\startdata
\multicolumn{5}{c}{Fit (a): $\chi^2_{\nu} = 1.51$, $rms=4.72\mps$}\\
\noalign{\vskip .50ex}
\cline{1-5}
\noalign{\vskip .50ex}
$K~ (\mps)$ && 95.0 && 63.0 \\
$P$ (days) && 450.5 && 436.7 \\
$e$ && 0.516 && 0.678 \\
$\omega$ (deg) && 126 && 135\\
$\mathcal{M}$ (deg) && 322 && 129 \\
$i$ (deg) && 110.0 &&10.4 \\
$\Omega$ (deg) && 0.0 && 327 \\
$m$ ($M_{\rm{J}}$)  && 3.58 && 10.47 \\
$V_1 ~(\mps)$ && \multicolumn{3}{c}{-8.7} \\
$V_2 ~(\mps)$ && \multicolumn{3}{c}{1.8}\\
\noalign{\vskip .50ex}
\cline{1-5}
\noalign{\vskip .50ex}
\multicolumn{5}{c}{Fit (b): $\chi^2_{\nu} = 1.61$, $rms=4.87\mps$}\\
\noalign{\vskip .50ex}
\cline{1-5}
\noalign{\vskip .50ex}
$K~ (\mps)$ && 88.0 && 64.5 \\
$P$ (days) && 446.4 && 439.9 \\
$e$ && 0.460 && 0.701 \\
$\omega$ (deg) && 111 && 130\\
$\mathcal{M}$ (deg) && 331 && 130 \\
$i$ (deg) && 88.1 && 8.1 \\
$\Omega$ (deg) && 0.0 && 165.1 \\
$m$ ($M_{\rm{J}}$)  && 3.22 && 13.31 \\
$V_1 ~(\mps)$ && \multicolumn{3}{c}{-9.1} \\
$V_2 ~(\mps)$ && \multicolumn{3}{c}{0.7}\\
\noalign{\vskip .50ex}
\cline{1-5}
\noalign{\vskip .50ex}
\multicolumn{5}{c}{Fit (c): $\chi^2_{\nu} = 1.59$, $rms=4.85\mps$}\\
\noalign{\vskip .50ex}
\cline{1-5}
\noalign{\vskip .50ex}
$K~ (\mps)$ && 89.0 && 65.9 \\
$P$ (days) && 455.3 && 442.0 \\
$e$ && 0.480 && 0.641 \\
$\omega$ (deg) && 135 && 156\\
$\mathcal{M}$ (deg) && 317 && 122 \\
$i$ (deg) && 32.8 && 3.2 \\
$\Omega$ (deg) && 0.0 && 161.1 \\
$m$ ($M_{\rm{J}}$)  && 6.07 && 37.07 \\
$V_1 ~(\mps)$ && \multicolumn{3}{c}{-2.8} \\
$V_2 ~(\mps)$ && \multicolumn{3}{c}{-7.5}\\
\noalign{\vskip .50ex}
\cline{1-5}
\noalign{\vskip .50ex}
\multicolumn{5}{c}{Fit (d): $\chi^2_{\nu} = 1.43$, $rms=4.60\mps$}\\
\noalign{\vskip .50ex}
\cline{1-5}
\noalign{\vskip .50ex}
$K~ (\mps)$ && 93.6 && 83.0 \\
$P$ (days) && 440.4 && 451.0 \\
$e$ && 0.508 && 0.776 \\
$\omega$ (deg) && 125 && 141\\
$\mathcal{M}$ (deg) && 318 && 131 \\
$i$ (deg) && 10.0 && 162.3 \\
$\Omega$ (deg) && 0.0 && 71 \\
$m$ ($M_{\rm{J}}$)  && 19.12 && 6.98 \\
$V_1 ~(\mps)$ && \multicolumn{3}{c}{-10.6} \\
$V_2 ~(\mps)$ && \multicolumn{3}{c}{1.1}\\

\enddata
\end{deluxetable}

%%%%%%%%%%%%%%%%%%%%%%%%%%%%%%%%

%%%%%%%%  table of errors of coplanar inclined 2:1 MMR best-fit  %%%%%%%%%%%%%%

 \begin{deluxetable}{lcccccccc}
 \tablecolumns{9}	
 \tablewidth{0pt}
 \tablecaption{Uncertainties of fitting parameters for the coplanar inclined 2:1 MMR best fit
 \label{tab:error}}
 \tablehead{
 \colhead{Parameter}&\colhead{}&\colhead{Value}&\colhead{}&\colhead{}&\colhead{}&\colhead{Uncertainties (1-$\sigma$)}&\colhead{}&\colhead{}\\
}
\startdata
&&&& $\sqrt{C_{ll}}$ & & Constant $\Delta\chi^2$ && Bootstrap\\
\noalign{\vskip .50ex}
\cline{1-9}
\noalign{\vskip .50ex}
$K_1$ ($\mps$) & & 54.4 & & $\pm~2.0 $  &&  $_{-1.8}^{+2.0}$ && $_{-2.5}^{+3.4}$ \\
$P_1$ (days) && 219.3  &&  $\pm~0.8$ && $_{-0.6}^{+1.0}$ && $_{-1.0}^{+2.2}$\\
$e_1$ && 0.425 &&   $\pm~0.018$ && $_{-0.016}^{+0.016}$ && $_{-0.020}^{+0.030}$ \\
$\omega_1$ (deg) && 133 && $\pm~3$ && $_{-3}^{+3}$ && $_{-5}^{+6}$\\
$\mathcal{M}_1$ (deg) && 256 && $\pm~6$ &&$_{-5}^{+6}$ && $_{-10}^{+7}$  \\
$K_2$ ($\mps$) && 39.8 && $\pm~1.3$ && $_{-1.2}^{+1.2}$ && $_{-1.3}^{+2.6}$  \\
$P_2$ (days) && 442.4 && $\pm~3.1$ && $_{-3.3}^{+2.2}$ && $_{-7.9}^{+2.3}$ \\
$e_2$ &&  0.203 && $\pm~0.052$ &&$_{-0.053}^{+0.045}$  &&  $_{-0.065}^{+0.070}$ \\
$\omega_2$ (deg) &&  107 && $\pm~8$ && $_{-7}^{+9}$ && $_{-10}^{+13}$\\
$\mathcal{M}_2$ (deg) && 333 && $\pm~8$ &&$_{-9}^{+7}$ && $_{-13}^{+10}$ \\
$i $ (deg) && 19.41  && $\pm~4.13$ && $_{-3.33}^{+4.42}$ && $_{-5.52}^{+4.86}$ \\
\\
$a_1$ (AU) & & 0.7423 & & && && $_{-0.0016}^{+0.0051}$ \\
$a_2$ (AU) & & 1.1866 & & & & & & $_{-0.0125}^{+0.0041}$ \\
$m_1$ ($M_{\rm{J}}$) & & 4.78 & & & & & & $_{-0.89}^{+1.78}$ \\
$m_2$ ($M_{\rm{J}}$) & & 4.80 & & & & & & $_{-0.88}^{+1.98}$ \\
\enddata
\tablecomments{Semimajor axis and planetary mass are not direct fitting parameters in our model, so their uncertainties here are only determined by the bootstrap method. The parameters listed here are suggested as the best-fit parameters for the HD 82943 planetary system, and the uncertainties determined by bootstrap method are suggested as the reported uncertainties.}
\end{deluxetable}

%%%%%%%%%% Table of MCMC and Bootstrap  %%%%%%%%%%%%%%
\begin{deluxetable}{lccccccccc}
\tabletypesize{\scriptsize}
%\rotate
\tablewidth{0pt}
 \tablecaption{Mean values and uncertainties of fitting parameters for the coplanar 2:1 MMR fits  from Bootstrap and MCMC 
 \label{tab:error4}}
 \tablehead{
 \colhead{Parameter}&\colhead{}&\colhead{}&\colhead{}&\colhead{}&\colhead{Mean $\pm 1\sigma$}&\colhead{}&\colhead{}&\colhead{}&\colhead{}\\
}
\startdata
&& \multicolumn{2}{c}{Keplerian Fits} && \multicolumn{2}{c}{$20^{\circ}$ Inclined Dynamical Fits } && \multicolumn{2}{c}{Edge-on Dynamical Fits}\\
\noalign{\vskip .50ex}
\cline{1-10}
\noalign{\vskip .50ex}
& & Bootstrap & MCMC && Bootstrap & DEMCMC && Bootstrap & DEMCMC \\
\noalign{\vskip .50ex}
\cline{1-10}
\noalign{\vskip .50ex}
$K_1$ ($\mps$)  & & $59.06^{+6.66}_{ -4.49}	$ & $60.04^{+3.11}_{ -3.10}	$  && $54.64^{+3.62}_{-2.32}$ & 	 $55.29^{+2.02}_{-1.92}$   && $57.67^{+3.18}_{ -3.65}	$ & $58.17^{+2.18}_{ -1.31}	$ \\
$P_1$ (days)   &&  $220.13^{+0.16}_{ -0.35}	$  & $220.17^{+0.12}_{ -0.12}	$ && $ 219.49^{+1.34}_{-0.67}$ &     $219.34^{+0.58}_{-0.49}$ &&$221.23^{+0.59}_{ -0.57}	$ & $220.98^{+0.22}_{ -0.17}	$  \\
$e_1$  &&  $0.420^{+0.035}_{ -0.042}	$ &  $0.402^{+0.024}_{ -0.022}	$  && $0.430^{+0.030}_{-0.019}$ &  $0.424^{+0.018}_{-0.018}$  && $0.410^{+0.031}_{ -0.019}	$ &  $0.406^{+0.010}_{ -0.012}	 $  \\
$\omega_1$ (deg)  && $115.3^{+12.3}_{ -5.1}	$ &  $118.8^{+5.5}_{ -5.8}	$ && $ 132.0^{+3.4}_{-3.2}$ &  $131.4^{+2.7}_{-2.6}$ &&  $122.4^{+2.9}_{ -3.0}$	&  $122.4^{+0.9}_{ -0.8}$  \\
$\mathcal{M}_1$ (deg)  && $270.5^{+4.3}_{ -14.1}	$ &  $270.3^{+3.9}_{ -3.9}	$ 	&& $255.3^{+4.8}_{-7.9}$ &  $258.3^{+4.6}_{-4.6}$ &&$277.1^{+5.1}_{ -6.8}	$ & $276.3^{+1.9}_{ -1.7}	$  \\
$a_1$ (AU) & & $0.744^{+0.0003}_{ -0.0009}	$ &  $0.744^{+0.0003}_{ -0.0003}	$  &&  $0.743^{+0.0030}_{-0.0015}$ &  $0.742^{+0.0013}_{-0.0011}$ && $0.746^{+0.0013}_{ -0.0013}	$  &   $0.745^{+0.0005}_{ -0.0004}	$    	 \\
$m_1 \sin(i)$ $(M_{\rm{J}})$ & & $1.729^{+0.239}_{-0.155}$ &  $1.771^{+0.104}_{-0.109}$ && $4. 653^{+0.308}_{-0.191}$ &  $4.724^{+0.429}_{-0.374}$ & & $1.70^{+0.105}_{-0.114}$ &  $1.714^{+0.076}_{-0.043}$  	\\
\\
$K_2$ ($\mps$) && $39.39^{+2.85}_{ -1.38}	$ &  $37.70^{+1.24}_{ -1.10}	$  && $40.39^{+2.13}_{-1.80} $& $39.72^{+1.29}_{-1.22}$   && $38.50^{+1.97}_{ -1.28}	$ &  $37.82^{+0.50}_{ -0.45}	$  \\
$P_2$ (days)  && $440.84^{+1.51}_{ -1.18}	$ &  $440.60^{+0.80}_{ -0.77}	$  && $ 441.75^{+2.68}_{-3.76}$ &  $442.48^{+1.89}_{-2.30}$ &&  $437.97^{+1.29}_{ -1.39}	$  &   $438.51^{+0.41}_{ -0.44}	$ \\
$e_2$  && $0.211^{+0.069}_{ -0.058}	$ &  $0.114^{+0.073}_{ -0.075}	$ && $ 0.210^{+0.055}_{-0.099} $ &  $0.183^{+0.048}_{-0.053} $&& $0.133^{+0.115}_{ -0.068}	$ & $0.087^{+0.038}_{ -0.053}	$  \\
$\omega_2$ (deg)  && $123.9^{}_{ -54.8}$  &   $94.7^{+211.2}_{ -58.7}$ && $111.1_{-9.4}$ &  $105.5^{+9.9}_{-8.6}$&& $114.3^{}_{ -20.8}	$ &  $89.3^{+7.5}_{ -53.3}	$   \\
$\mathcal{M}_2$ (deg)  && $152.4^{+202.7}_{ -118.2}	$ & $108.7^{+222.4}_{ -84.0}	$ && $ 327.1^{+9.9}$ &  $333.9^{+8.1}_{-9.8}$&&$246.7^{+47.4}_{ }$ & $338.7^{+8.4}_{ -281.9}$   \\
$a_2$  (AU) & & $1.182^{+0.003}_{ -0.002}	$  &  $1.181^{+0.001}_{ -0.001}	$ && $1.185^{+0.0048}_{-0.0067}$ &  $1.187^{+0.0034}_{-0.0041}$  & & $1.1767^{+0.0023}_{ -0.0025}$ &  $1.1780^{+0.0007}_{ -0.0007}$ 	\\
$m_2 \sin(i)$ $(M_{\rm{J}})$ & & $1.561^{+0.067}_{ -0.049}$ &  $1.518^{+0.042}_{ -0.041}$ & & $ 4.70^{+0.175}_{-0.159}$ &  $4.643^{+0.416}_{-0.339}$  & & $1.544^{+0.050}_{ -0.046}$ &  $1.527^{+0.015}_{ -0.016}$  \\
%\\
%$V_1 $ $(\mps)$ & &  $-5.84^{+3.04}_{ -1.81}$	&  $-6.26^{+1.44}_{ -1.43}$ && $ & &  $-8.12^{+2.34}_{ -2.06}$ &  $-8.0^{+0.61}_{ -0.67}$  \\
%$V_2 $ $(\mps)$ & &  $-8.04^{+1.11}_{ -0.93}$ &  $-8.20^{+0.85}_{ -0.86}$	 & &  $-8.32^{+0.85}_{ -0.85}$	&  $-8.24^{+0.26}_{ -0.31}$ & & $-8.29^{+0.84}_{ -0.84}$	 & $-8.25^{+0.27}_{ -0.31}$

\enddata
\tablecomments{Blank in some errors means that the error cannot be determined.}
\end{deluxetable}
%%%%%%%%%%%%% end table  % %%%%%%%%%%%%%%%%

\clearpage

%%%%%%%%%%%%%%%%%%%%%%%%%%%%%%%%%%%%%
%%%%%%%%%%%%%  FIGURE %%%%%%%%%%%%%%%%%%%%

%%%%%%%%  figure of edge-on RV curve and residuals  %%%%%%%%%%%%
%%%%%%%%%%%%%%%%%%%%%%%%%%%%%%%%%%%%%%%%%

%%%%%%%  figure of chisq contours  at edge on orbits  %%%%%%%%%%%%%%%
%%%%%%%%%%%%%%%%%%%%%%%%%%%%%%%%

%%%%%%%%%% figure of dynamical properties of edge-on orbits %%%%%%%%%%

%%%%%%%%%%%%%%%%%%%%%%%%%%%%%%%%

%%%%%%%%%%%  figure of dynamical properties of edge-on orbits %%%%%%%%%%%

%%%%%%%%%%%%%%%%%%%%%%%%%%%%%%%%

%%%%%%% figure of chisq changes as function of inclination %%%%%
%%%%%%%%%%%%%%%%%%%%%%%%%%%%%%%%%%%%
\clearpage

%%%%%%figure of para changes with inclination %%%%%%%%%
%%%%%%%%%%%%%%%%%%%%%%%%%%%%%%%%%%%%

%%%%%%%%  figure of 20 inclined RV curve and residuals  %%%%%%%%%%%%
%%%%%%%%%%%%%%%%%%%%%%%%%%%%%%%%%%%%%%%%%

%%%%%%%%%% figure of chisq contours of ifree orbits %%%%%%%%%%

%%%%%%%%%%%%%%%%%%%%%%%%%%%%%%%%

%%%%%%%%%% figure of discontinued area %%%%%%%%%%%%%%%%%%%%%%
%%%%%%%%%%%%%%%%%%%%%%%%%%%%%%%%%%%%%%%%%%%%%%

%%%%%%%%%% figure of chisq properties of 21ifree orbits %%%%%%%%%%

%%%%%%%%%%%%%%%%%%%%%%%%%%%%%%%%

%%%%%%%%%%%  figure of dynamical properties of  i free orbits %%%%%%%%%%%

%%%%%%%%%%%%%%%%%%%%%%%%%%%%%%%%

%%%%%%%%%%%  figure of dynamical properties mutually inclined 2:1MMR orbits %%%%%%%%%%%
\clearpage

%%%%%%%%%%%%%%%%%%%%%%%%%%%%%%%%

%%%%%%% figure of periodogram %%%%%%%%%%

%%%%%%%%%%%%%%%%%%%%%%%%%%%%%

%%%%%%%%%  figure of 3-planet dynamical fit %%%%%%%%%%

 %%%%%%%%%%%%%%%

%%%%%%%%%  figure of 1:1 resonance dynamical fit %%%%%%%%%%

 %%%%%%%%%%%%%%%
 
 %%%%%%%%%  figure of 11 dynamical fit %%%%%%%%%%

 %%%%%%%%%%%%%%%
 
 \clearpage
 
 %%%%%%%%%  figure posteriori of coplanar dynamical fit from bootstrap %%%%%%%%%%

 %%%%%%%%%%%%%%%

%%%%%%%%%%%%%%%%%%%%%%%%%%%%%%%%%%%%%%%%%%%%%%%%%%

%%%%%%%%  Figure of Histogram of Bootstrap and MCMC (Kepfit) %%%%%
%%%%%%%%%%%  end figure %%%%%%%%%
\clearpage

%%%%%%%%  Figure of Histogram of Bootstrap and DEMCMC edge-on fits %%%%%
%%%%%%%%%%%  end figure %%%%%%%%%

 %%%%%%%%%%%%%%%
 
\end{document}